\documentclass[aps,prl,twocolumn,superscriptaddress,showpacs,altaffillsymbol,nofootinbib]{revtex4-1} 
\usepackage{amssymb}
\usepackage{amsmath}
\usepackage{commath}
\usepackage{graphicx}
\usepackage{verbatim}
\usepackage{epsfig}
\usepackage{float}
\usepackage{epstopdf}
\usepackage{color}
\usepackage{siunitx}
\usepackage{array}
\usepackage{physics}
\usepackage{gensymb}
\newcolumntype{P}[1]{>{\centering\arraybackslash}p{#1}}
\newcolumntype{M}[1]{>{\centering\arraybackslash}m{#1}}

\usepackage{hyperref}
\hypersetup{breaklinks=true, colorlinks=true, allcolors = blue}

\usepackage{xr-hyper}
\makeatletter
\newcommand*{\addFileDependency}[1]{
  \typeout{(#1)}
  \@addtofilelist{#1}
  \IfFileExists{#1}{}{\typeout{No file #1.}}
}
\makeatother

\newcommand*{\myexternaldocument}[1]{%
    \externaldocument{#1}%
    \addFileDependency{#1.tex}%
    \addFileDependency{#1.aux}%
}

\myexternaldocument{SM}

\begin{document}

\title{Evidence for bootstrap percolation dynamics in a photo-induced phase transition}

\author{Tyler Carbin}
\email{tcarbin@g.ucla.edu}
\affiliation{Department of Physics and Astronomy, University of California Los Angeles, Los Angeles, CA 90095-1547}

\author{Xinshu Zhang}
\affiliation{Department of Physics and Astronomy, University of California Los Angeles, Los Angeles, CA 90095-1547}

\author{Adrian B. Culver}
\affiliation{Department of Physics and Astronomy, University of California Los Angeles, Los Angeles, CA 90095-1547}
\affiliation{Mani L. Bhaumik Institute for Theoretical Physics, Department of Physics and Astronomy, University of California Los Angeles, Los Angeles, CA 90095}

\author{Hengdi Zhao}
\affiliation{Department of Physics, University of Colorado at Boulder, Boulder, Colorado 80309, USA}

\author{Alfred Zong}
\affiliation{Department of Chemistry, University of California at Berkeley, Berkeley, CA, 94720, USA}
\affiliation{Materials Sciences Division, Lawrence Berkeley National Laboratory, Berkeley, CA, 94720, USA}

\author{Rishi Acharya}
\affiliation{Department of Physics and Astronomy, University of California Los Angeles, Los Angeles, CA 90095-1547}

\author{Cecilia J. Abbamonte}
\affiliation{Department of Physics and Astronomy, University of California Los Angeles, Los Angeles, CA 90095-1547}

\author{Rahul Roy}
\affiliation{Department of Physics and Astronomy, University of California Los Angeles, Los Angeles, CA 90095-1547}
\affiliation{Mani L. Bhaumik Institute for Theoretical Physics, Department of Physics and Astronomy, University of California Los Angeles, Los Angeles, CA 90095}

\author{Gang Cao}
\affiliation{Department of Physics, University of Colorado at Boulder, Boulder, Colorado 80309, USA}

\author{Anshul Kogar}
\email{anshulkogar@physics.ucla.edu}
\affiliation{Department of Physics and Astronomy, University of California Los Angeles, Los Angeles, CA 90095-1547}

\date{\today}

\begin{abstract}

Upon intense femtosecond photo-excitation, a many-body system can undergo a phase transition through a non-equilibrium route, but understanding these pathways remains an outstanding challenge. Here, we use time-resolved second harmonic generation to investigate a photo-induced phase transition in $\mathrm{Ca}_3\mathrm{Ru}_2\mathrm{O}_7$ and show that mesoscale inhomogeneity profoundly influences the transition dynamics.  
We observe a marked slowing down of the characteristic time, $\tau$, that quantifies the transition between two structures. $\tau$ evolves non-monotonically as a function of photo-excitation fluence, rising from below 200~fs to $\sim$1.4~ps, then falling again to below 200~fs. To account for the observed behavior, we perform a bootstrap percolation simulation that demonstrates how local structural interactions govern the transition kinetics. Our work highlights the importance of percolating mesoscale inhomogeneity in the dynamics of photo-induced phase transitions and provides a model that may be useful for understanding such transitions more broadly. 

\end{abstract}

\maketitle

In a photo-induced phase transition (PIPT), a qualitative and macroscopic change to the behavior of a many-body system occurs following intense femtosecond photo-excitation. PIPTs are inherently different from equilibrium transitions because they typically proceed through a far from equilibrium, non-thermal pathway where time becomes a fundamental variable.
Tracking the temporal evolution of the spatially averaged response functions is crucial to understanding the spectacular behaviors instigated by photo-excitation in solids, such as the appearance of transient order and metastability of hidden states~\cite{Stojchevska2014, Zhang2016,Mitrano2016,Fausti2011,Kogar2020,McIver2020}. However, order often evolves in a spatially non-uniform manner in many PIPTs; a major recurring theme is the presence of electronic and crystallographic inhomogeneity on the nano- to micro-scale~\cite{Perez-Salinas2022,Mihailovic2005,Abreu2015,Lee2012,Abreu2020,Wall18,Sternbach21,Johnson22,OCallahan2015,Donges2016,Hilton2007,Otto18,Madaras20,Rodriguez-Vega15,Cocker2012,Haupt2016,Laulhe17,Danz2021,Zong2019}.

Experimentally capturing the \textit{dynamics} of mesoscale structures has proven difficult.
However, inhomogeneous textures have been observed in materials exhibiting long-lived metastability following photo-excitation. In the metastable ``hidden" states of both 1\textit{T}-Ta$\mathrm{S}_2$~\cite{Stojchevska2014,Cho2016, Ma2016} and strained $\mathrm{La}_{2/3}\mathrm{Ca}_{1/3}\mathrm{MnO}_3$~\cite{Mcleod2020, Zhang2016}, quasi-static textures were observed using real-space scanning probe techniques.
But, such experimental approaches are currently unfeasible for observing inhomogeneity that evolves on the short timescales characteristic of many PIPTs.
A notable exception is the insulator-metal PIPT in $\mathrm{VO}_2$, which was found to exhibit similar transient textures determined by grain boundaries or pre-existing domains following each applied laser pulse~\cite{Sternbach21,Johnson22,Donges2016}.
To understand the effects of the dynamic inhomogeneity, our approach here is to quantify its aggregate effects on macroscopic observables and correlate the observations with a percolation model. 

To accomplish this goal, we employ time-resolved second harmonic generation (SHG) to investigate a PIPT in a prototypical correlated material, Ca$_3$Ru$_2$O$_7$ (CRO), in which photo-induced inhomogeneity is expected to occur (see below). By relating the experimental observations to simulation results, we provide strong evidence that structural percolation, mediated by lattice strain, governs the transition kinetics. Specifically, we show that the photo-induced dynamics are consistent with bootstrap percolation (BP), a particular cellular automata model that lacks detailed balance.

\begin{figure}
    \centering
    \includegraphics[scale=0.33]{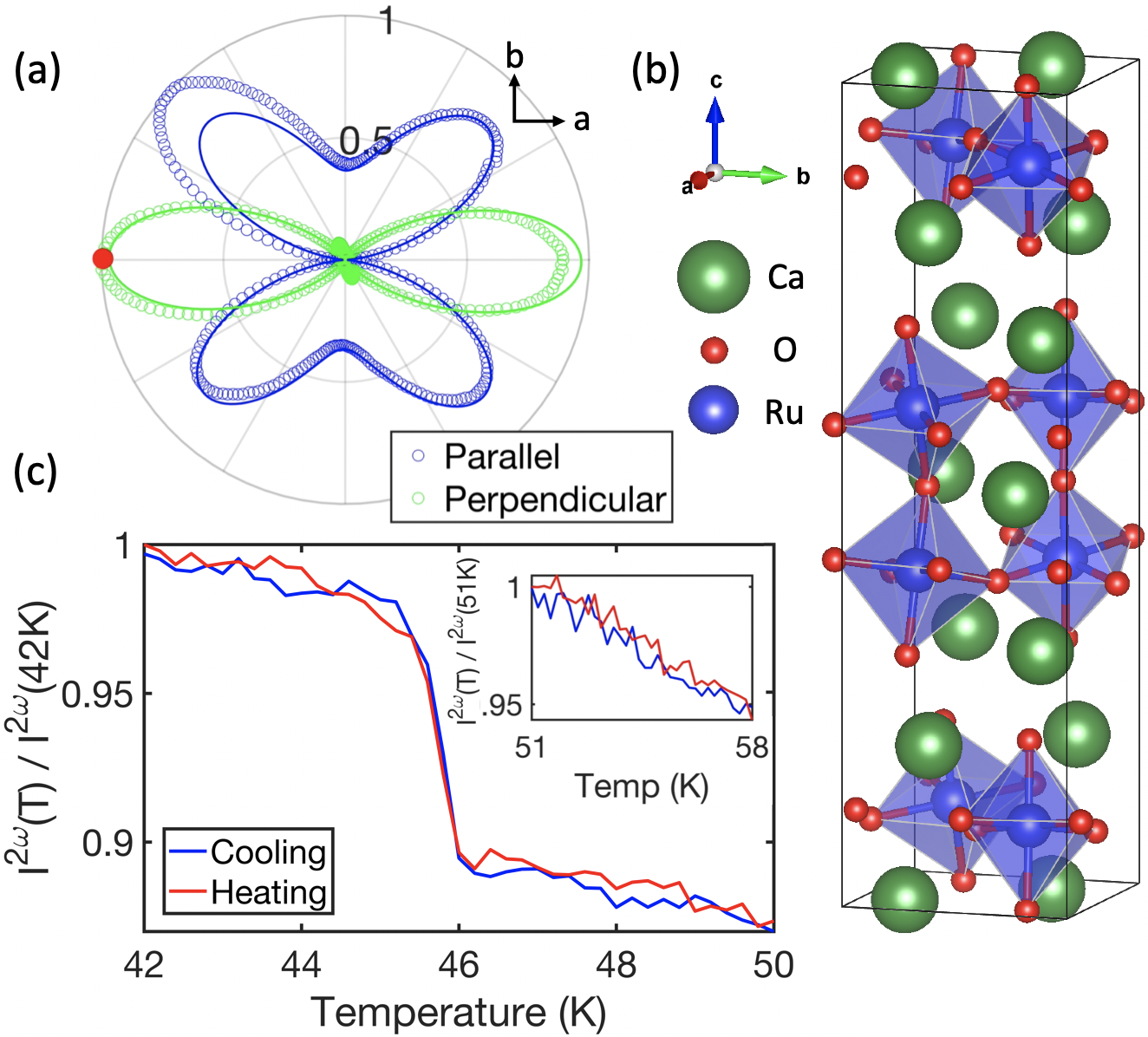}
    \caption{\textbf{(a)} Measured RA-SHG patterns at 52~K with incident and outgoing polarizers in parallel and perpendicular geometry. Simultaneous fits to both channels are obtained using a susceptibility tensor constrained by the $C_{2v}$ point group (solid lines). The data are normalized to a maximum in the perpendicular channel. The $a$ and $b$ crystallographic axes are indicated with black arrows. \textbf{(b)} Schematic of the low-temperature crystal structure. \textbf{(c)} Temperature dependence of the SHG intensity $I^{2\omega}$ at a polarization angle indicated by the red dot in \textbf{(a)}. In this geometry, $I^{2\omega}\propto|\chi^{ED}_{baa}|^2$. The red(blue) curve corresponds to heating(cooling). (Inset) Normalized $I^{2\omega}\propto|\chi^{ED}_{baa}|^2$ across $T_N$.}
    \label{Fig1}
\end{figure}

$\mathrm{Ca}_3\mathrm{Ru}_2\mathrm{O}_7$ (CRO) 
is the $n$=2 compound in the Ruddlesden-Popper series $\mathrm{Ca}_{n+1}\mathrm{Ru}_{n}\mathrm{O}_{3n+1}$~\cite{Yoshida05}. 
The crystal is distorted from the typical \textit{I}4\textit{/mmm} structure due to rotation and tilts of the oxygen octahedra around the (001) axis and (110) axis, respectively~(Fig.~\ref{Fig1}(b)).  
At all temperatures, its crystallographic space group is $Bb2_{1}m$ (No. 36), with point group $C_{2v}$.
As the temperature is reduced below the Néel temperature, $T_{N}$=56~K, CRO undergoes a continuous phase transition to an antiferromagnetic state in which the Ru spins are aligned ferromagnetically along the $\pm a$ axis within each bilayer and antiferromagnetically between bilayers~\cite{Bao2008}. 
Of primary interest here is the discontinuous metal-insulator transition at $T_{MI}$=48~K. In samples grown by the floating zone method, the low-temperature state is semimetallic~\cite{pnas.2003671117, Horio21,Baumberger2006}, but in the flux-grown samples used here, this state is truly insulating~\cite{Cao97}. This electronic transition is accompanied by a rotation of the spins from the $\pm a$ axis to the $\pm b$ axis~\cite{Bao2008,Bohnenbuck08} and a structural transition without a change in crystallographic symmetry. Through the transition, the $c$ axis contracts by $\sim$0.1\% and the $a$ and $b$ axis lattice parameters expand by $\sim$0.07\%~\cite{Yoshida05}. The structural change leads to a compression of the oxygen octahedra and further breaks the degeneracy between the Ru $d_{xy}$ and the $d_{xz/yz}$ orbitals; it is therefore thought to be a vital component of the insulator-metal transition by promoting an orbital polarization~\cite{Karpus04,Peng2017,Lechermann2020,Bohnenbuck08,Puggioni20}.

To probe the dynamics of the phase transition, we perform time-resolved SHG, a technique that can monitor the symmetry of CRO in its various phases~\cite{FiebigReview}.
These measurements were performed in a reflection geometry, using 180~fs laser pulses with a 5~kHz repetition rate.  We collected data in two configurations -- one with parallel and one with perpendicular incident and outgoing light polarization. The probe beam was centered at 900~nm (1.38~eV) and was shone normal to the $ab$-plane with a 40~$\mu$m spot size. The pump beam was centered at 1030~nm (1.20~eV) and was incident at 15$\degree$ to the surface normal with a 200~$\mu$m spot size. When the system is pumped with the laser pulse, the excited electrons remain within the $t_{2g}$ manifold of the Ru$^{4+}$ atoms (the crystal field splitting to the $e_g$ levels is $\sim$2~eV)~\cite{Bertinshaw21}.

\begin{figure}
    \centering
    \includegraphics[scale=0.64]{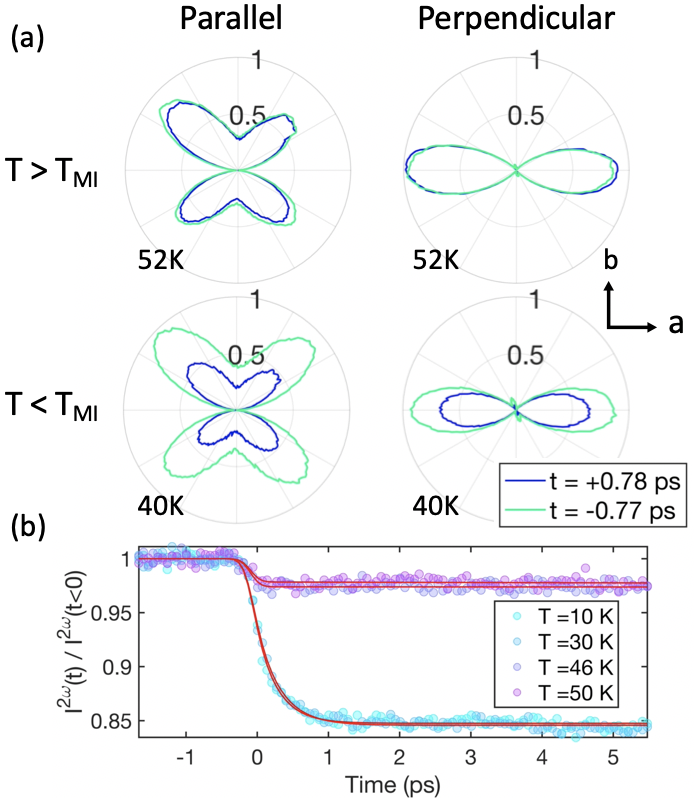}
    \caption{Temperature-dependent SHG response to the 1030~nm pump pulse. \textbf{(a)} Rotational anisotropy patterns measured before (t$<$0) and after (t$>$0) a  0.28 mJ/$\mathrm{cm}^2$ pump pulse both above and below $T_{MI}$. The pulse induces a large drop in SHG intensity when applied below $T_{MI}$ and has little effect above $T_{MI}$. \textbf{(b)} Time traces of the normalized SHG intensity $I^{2\omega}\propto |\chi^{ED}_{baa}|^2$ at several temperatures with a 0.86~mJ/$\mathrm{cm}^2$ pulse. Red lines are fits to Eq. \eqref{Fpumpprobe}. Traces at additional temperatures are excluded here for clarity and are presented in the Supplemental Material~\cite{Supp}.}
    \label{Fig2}
\end{figure}

In the equilibrium state, the leading-order electric-dipole (ED) SHG contribution is allowed in CRO due to broken inversion symmetry (Fig.~\ref{Fig1}). At all temperatures, we obtain a good simultaneous fit to the parallel and perpendicular polarization configurations of the rotational anisotropy (RA) pattern with the ED contribution of the known point group symmetry $C_{2v}$~\cite{Supp} (Fig. \ref{Fig1}(a)). In Fig.~\ref{Fig1}(c), we show the temperature dependence of the SHG intensity, $I^{2\omega}$, arising from the tensor element $\chi^{ED}_{baa}$ (red dot in Fig.~\ref{Fig1}(a)) across the insulator-metal transition. Although there is no temperature-induced change in symmetry of the RA pattern across the insulator-metal transition (Fig.~\ref{Fig2}(a)), the intensity exhibits a pronounced jump. (The offset between the reported value for $T_{MI}$=48~K and the observed jump around 46~K arises due to laser heating)~\cite{Supp}. No thermal hysteresis is measured, and no features are observed at $T_N$ (Fig.\ref{Fig1}(c) inset).

Because the crystal, electronic and magnetic structure all change at $T_{MI}$, the cause of the observed jump in $I^{2\omega}$ is not immediately clear. A previous study reports a similar increase in $I^{2\omega}$ for pure CRO, but observes no such feature in the Fe-doped compound $\mathrm{Ca}_{3}\mathrm{Ru}_{1.95}\mathrm{Fe}_{0.05}\mathrm{O}_7$~\cite{Lei19}. This difference is striking because, similar to pure CRO, the latter compound undergoes a transition in which the spins reorient from pointing along the $\pm a$ axis to the $\pm b$ axis and a concomitant metal-insulator transition.
However, unlike in pure CRO, this transition is not accompanied by a large structural change~\cite{Ke2014,Peng2017}. We therefore conclude that the increase in $I^{2\omega}$ is indicative of the structural change (note that this is consistent with no features being observed at $T_N$ Fig.~\ref{Fig1}(c)). To corroborate the connection between crystal structure and $I^{2\omega}$, in the Supplemental Material we use a Landau theory approach to show that a first-order transition with a symmetry-preserving order parameter can give rise to a jump in $I^{2\omega}$~\cite{Supp}.

We now study the PIPT instigated by intense femtosecond laser pulses. 
Figure \ref{Fig2}(a) shows the change in the RA-SHG pattern above and below $T_{MI}$ after the arrival of a 0.28~mJ/$\mathrm{cm}^2$ pump pulse. Below $T_{MI}$, there is a clear drop in intensity following the pulse, while the pattern above $T_{MI}$ is minimally affected. 
As in the thermal transition, the symmetry is unchanged. 
Figure~\ref{Fig2}(b) shows the evolution of $I^{2\omega} \propto |\chi^{ED}_{baa}|^2$ after photo-excitation at several temperatures. These curves demonstrate that the decrease in intensity is stable for $\gg5$ ps. 
The magnitude of the drop in $I^{2\omega}$ is roughly constant for various temperatures below $T_{MI}$, but is markedly smaller in the high-temperature phase. 
(Here the measured $T_{MI}$ is between 44-45 K due to laser heating from both pump and probe pulses). It should be noted that the intensity of the second harmonic,  $I^{2\omega}$, varies across the sample surface (the $ab$ plane) due to the nonuniform distribution of 180$\degree$ polar domains, as discussed more thoroughly in Refs.~\cite{Lei, Supp}. However, the photo-induced changes are associated only with the phase transition~\cite{Supp}.

\begin{figure}
    \centering
    \includegraphics[scale=0.6]{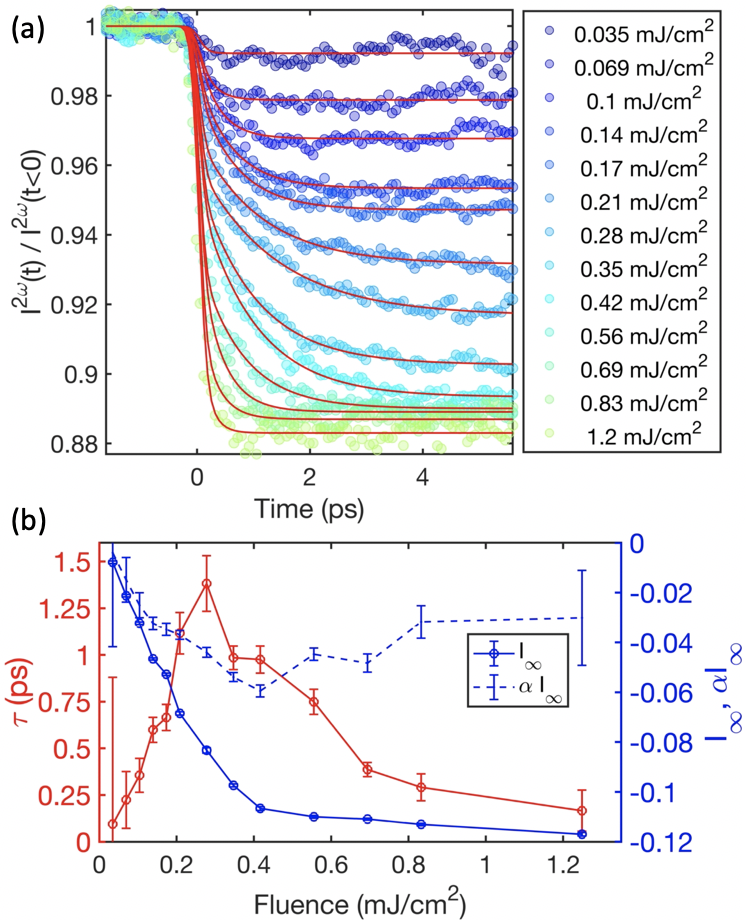}
    \caption{\textbf{(a)} Time evolution of the normalized SHG intensity, $I^{2\omega} \propto |\chi^{ED}_{baa}|^2$, for varying pump fluences, measured at a nominal temperature of 4~K (laser heating raises the temperature). Fits to Eq. \eqref{Fpumpprobe} are overlaid in red. \textbf{(b)} Best-fit values of $I_\infty$, $\alpha I_\infty$, and $\tau$ plotted vs. pump fluence. Error bars are 95\% confidence intervals from the fitting procedure.}
    \label{Fig3}
\end{figure}

To understand the kinetics of the PIPT, we measure SHG time traces in the low temperature state for various pump fluences (Fig. \ref{Fig3}(a)). 
We fit each time trace to the phenomenological function~\cite{Supp}:
\begin{equation}
    u(t)=1+\Bigl[\theta(t-t_0)\\
     I_{\infty}(1-\alpha e^{-(t-t_0)/\tau})\Bigr] \ast g(w_0,t),
     \label{Fpumpprobe}
\end{equation}
where $\theta(t)$ denotes the Heaviside step function, $g(w_0,t)$ is the cross-correlation of the pump and probe pulses and the $\ast$ indicates convolution.
This allows us to extract three parameters: (i) the time constant of the transient decay, $\tau$; (ii) the SHG intensity at late times relative to the intensity before the pulse, $I_{\infty}$; and (iii) the fraction of $I_\infty$ that is related to the dynamics associated with $\tau$, $\alpha$.
The best-fit values for $\tau$, $I_{\infty}$, and $\alpha I_\infty$ are plotted as a function of fluence in Fig. \ref{Fig3}(b). 
We find that $I_{\infty}$ decreases with increasing fluence until reaching a saturation $I_{\infty}^{sat}\approx$ -0.11 at fluence $F_{sat}\approx~0.4$~mJ/$\mathrm{cm}^2$, while $\alpha I_\infty$ decreases from zero to roughly half of $I_{\infty}^{sat}$ near $F_{sat}$ before increasing at high fluence.
The time constant $\tau$ exhibits the most noteworthy behavior; it first increases by nearly an order of magnitude from $<$200 fs to $\sim$1.4~ps, peaking near $F_{sat}$ before decreasing to roughly 200~fs.

For fluences $F\gtrsim F_{sat}$, these measurements suggest that the sample has reverted to the high-temperature structure; the intensity jump observed as a function of temperature (Fig.~\ref{Fig1}(c)) is completely suppressed by the photo-exciting laser pulse. However, the nature of the ``intermediate" states, characterized by $0<|I_{\infty}|<|I_{\infty}^{sat}|$ is not immediately obvious.
There are two natural possible scenarios to attribute to these states. 
In the first scenario, the lattice parameters would change in a spatially uniform manner throughout the illuminated region and would take on a value between that of the low- and high-temperature equilibrium phases.
Alternatively, the effect of the pulse could be spatially nonuniform and the intermediate states could consist of small regions in which the lattice parameters primarily take on either their low- or high-temperature equilibrium values. 

For several reasons, including the discontinuous character of the equilibrium transition and the observation of structural inhomogeneity in the hysteresis region of Ti-substituted CRO~\cite{McLeod21, Rabinovich2022}, we expect \textit{a priori} that the lattice parameters exhibit discontinuous changes. We show below that this scenario accounts for the experimental observations, most notably the non-monotonic fluence-dependence of the timescale $\tau$ (Fig. \ref{Fig3}(b)). 

To demonstrate that binary values of the lattice parameters can give rise to this behavior, we perform a bootstrap percolation simulation.
\begin{figure}
    \centering
    \includegraphics[scale=0.61]{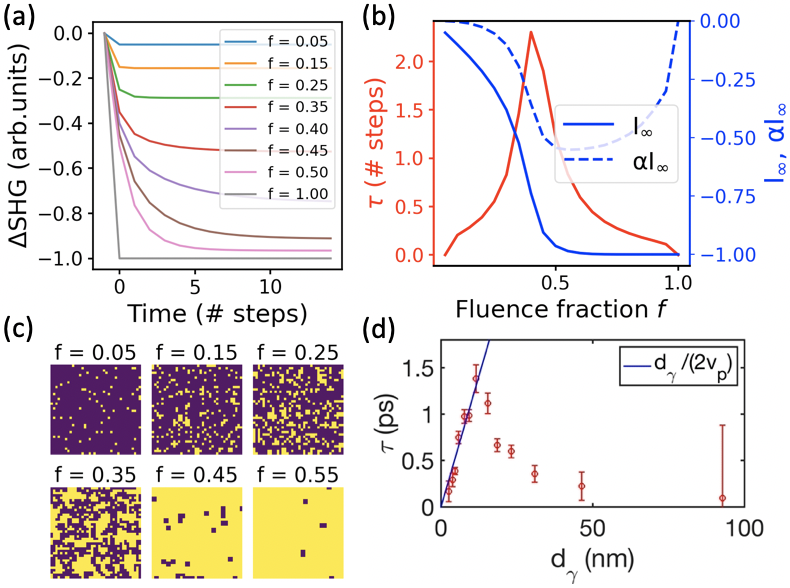}
    \caption{\textbf{(a)} Simulated temporal evolution of SHG intensity for various fluence fractions $f$ with strain threshold $\sigma_{th}$=3. $\Delta$SHG is determined from the fraction of sites in the $L_c$ state. Time steps indicate iterations of the simulation. (When compared to the experiment, each time step corresponds to a few hundred fs.) 
    \textbf{(b)} The parameters $\tau$, $I_{\infty}$, and $\alpha I_{\infty}$ determined from the simulations plotted against the fluence fraction parameter $f$. \textbf{(c)} Examples of two-dimensional slices of the final system state for various fluence fractions $f$. $L_c$($S_c$) sites are depicted in yellow(purple). We estimate that each site corresponds to a region of the material with a length scale on the order of ~1 nm~\cite{Supp}. \textbf{(d)} $\tau$ plotted versus the average distance between absorbed pump photons $d_\gamma$. The linear fit to the data in the high-fluence regime allows us to extract an approximate percolation speed $v_p$. }
    \label{Fig4}
\end{figure}
The simulation consists of the following recipe~\cite{Adler91}. (In the following, for ease of presentation, we refer only to the $c$-axis changes, but this is meant to represent the changes to all lattice parameters.) (1) An array of sites is defined that can take only a long or short $c$-axis lattice parameter, denoted $L_c$ and $S_c$, respectively. Before $t = 0$ the system is initialized to possess only $S_c$ sites, corresponding to the low-temperature insulating state. (2) At $t = 0$, a random subset of sites is switched to the $L_c$ state to mimic the effect of the pump. The number of switched sites is assumed to be proportional to the incident pump fluence. (3) Each remaining $S_c$ site then evolves according to the governing rule that if the number of nearest neighbor $L_c$ sites \textit{exceeds} a threshold value, $\sigma_{th}$, the examined site switches from $S_c$ to $L_c$. (4) Lastly, once converted to an $L_c$ site, it is forbidden from reverting to an $S_c$ one. These rules encompass the entire simulation, and it is run in discrete time steps until the system reaches quasi-equilibrium where site-switching no longer occurs. Imposition of the rule (4) is motivated by data showing that the recovery to the state with a globally contracted $c$-axis occurs on much longer timescales~\cite{Supp}. Because we disallow $L_c$-to-$S_c$ conversion, the model describes a manifestly non-equilibrium process characterized by a breakdown of detailed balance and a transition to an absorbing state. With this minimal model taking a single input parameter, $\sigma_{th}$, we are able to capture the qualitative behavior of all three fitted parameters in our data.

In our implementation, the sample is modeled as a 40$\times$40$\times$40 cubic array of sites.
The fraction of sites that are excited at $t$=0 is given by the fluence fraction parameter $f$, which is defined between 0 and 1, corresponding, respectively, to no pump pulse and to a pulse that excites all of the sites quasi-instantaneously. 
At each time step, the ``strain" at each site, $\sigma$, is equal to the number of neighboring sites in the $L_c$ state~\cite{Supp}. In the results of the simulation shown in Fig.~\ref{Fig4}, the threshold parameter, $\sigma_{th}$, is equal to three. To correlate the simulation to our data, we make the assumption that the change in $I^{2\omega}$ at each time step is proportional to the fraction of sites that have switched to the $L_c$ state. This correspondence allows us to produce the simulated time traces in Fig.~\ref{Fig4}(a). 
Finally, we fit the simulated time traces to Eq.~\eqref{Fpumpprobe} (without the finite pulse width factor $g(w_0,t)$) to extract $I_{\infty}$, $\alpha$, and $\tau$ as a function of the fluence fraction $f$ (Fig.~\ref{Fig4}(b))~\cite{Supp}.

Figure 4 shows that this model reproduces all of the qualitative aspects of the SHG response.
We emphasize that these simulations did not require fine-tuning to produce the desired results. 
Rather, we find that the non-monotonic fluence dependences of $\tau$ and of $\alpha$, with peaks near $F_{sat}$, are robust to changes of $\sigma_{th}$ and to the inclusion of additional neighbor couplings, both when the simulation is confined to two dimensions and when the penetration depths of the pump and probe pulses are taken into account.
See the Supplemental Material~\cite{Supp} for details of the simulations using alternate settings. 

We are left with the following physical picture to explain the non-monotonic behavior of the timescale $\tau$. 
At low fluences, random isolated sites are photo-excited to the $L_c$ state, but these sites cannot percolate very far; the $S_c$ sites do not possess a sufficient number of $L_c$ nearest neighbors to trigger a switching event. The observed transition time is therefore on the order of the exciting laser pulse in the experiment. As the fluence is increased, some sites that remained in the $S_c$ state after the initial photo-excitation exceed the nearest neighbor strain threshold and turn into an $L_c$ site. The $L_c$ sites then start to percolate; switching events are able to trigger further switching events. Near the percolation threshold, where almost all the sites are switched, the transition time starts to lengthen considerably, mimicking critical slowing down~\cite{ZongDSD}. 
At higher fluences, the number of sites that switch to the $L_c$ state instantaneously is large and only a short time is needed to switch the remaining $S_c$ sites. 
Within this framework, the peak in $\tau$ physically represents the maximum time it takes for the $L_c$ site percolation to occur following the initial photo-excitation.

This physical picture also lends itself to a natural interpretation of the parameters $I_\infty$ and $\alpha$. In this scheme, $I_{\infty}$ represents the total number of switched sites, including both the quasi-instantaneous effects from photo-excitation and the subsequent percolation. On the other hand, $\alpha$ describes the fraction of sites that convert from $S_c$ to $L_c$ solely due to percolation, and its dynamics are associated with the timescale $\tau$. The dip in $\alpha I_\infty$ with varying fluence thus indicates that the volume of the sample induced to become $L_c$ through site-to-site spreading is largest at fluences near $F_{sat}$, in accordance to what would be expected near a percolation threshold (Fig.~\ref{Fig4}(b)). 
The interpretation of these parameters allows us to understand the PIPT as a percolation phenomenon mediated by local interactions between neighboring sites.

A physical justification that the interactions are mediated by lattice strain is obtained by an estimate of the percolation speed $v_p$. We first convert the fluence to an average distance between absorbed pump photons $d_\gamma$~\cite{Supp}. For fluences $F>F_{sat}$, this quantity characterizes the length over which an average $L_c$ region percolates. (This is not the case for $F<F_{sat}$ when $S_c$ regions will persist between sites excited by photons). We find that $\tau$ is linearly proportional to $d_\gamma$ in this high-fluence (low $d_\gamma$) regime, indicating that the percolation speed is independent of fluence. Performing a linear fit, we extract a characteristic growth speed of $\sim$4400~m/s (Fig.~\ref{Fig4}(d)). Though the speed of sound has not been measured in Ca$_3$Ru$_2$O$_7$, this value is in accord with what would be expected if the growth of $L_c$ clusters was given by ballistic strain propagation. 

In summary, our study provides substantial evidence that the kinetics of the PIPT in Ca$_3$Ru$_2$O$_7$ proceeds through the percolation of nanoscale clusters which is mediated by lattice strain. Specifically, the transition dynamics are qualitatively captured by a model of bootstrap percolation.
The simplicity of this model suggests that it may hold significant promise for understanding the dynamics of other PIPTs.
Indeed, time-resolved measurements of the photo-induced transition in VO$_2$~\cite{Wall13,Cocker2012,Liu22,Otto18} shows two timescales of comparable duration to those observed in this work, which may also be described within a percolation theory. 
Our work paves the way towards understanding the effects of dynamic inhomogeneity on PIPTs and provides a general model that may be useful for investigating photo-excited materials more broadly.

\vspace{1em}

\begin{acknowledgements}
We thank R. Schonmann for helpful discussions regarding the percolation model, M. Rasiah for help with the initial construction of the second harmonic generation setup, and S. Kivelson for helpful suggestions regarding the Landau theory calculation. Research at UCLA was supported by the U.S. Department of Energy (DOE), Office of Science, Office of Basic Energy Sciences under Award No. DE-SC0023017 (experiment and simulations). Work at UC Boulder was supported by the National Science Foundation via Grant No. DMR~2204811 (materials synthesis). A.Z. acknowledges support by the Miller Institute for Basic Research in Science. A.K., R.A. and C.J.A. acknowledge the REU program through STROBE: a National Science Foundation Science and Technology Center under award no. DMR-1548924.
\end{acknowledgements}

\bibliography{references}

\clearpage
\end{document}



\newcommand{\beq}{\begin{equation}}
\newcommand{\eeq}{\end{equation}}
\newcommand{\bseq}{\begin{subequations}}
\newcommand{\eseq}{\end{subequations}}
\newcommand{\bpmat}{\begin{pmatrix}}
\newcommand{\epmat}{\end{pmatrix}}
\newcommand{\bpl}{\boldsymbol{(}}
\newcommand{\bpr}{\boldsymbol{)}}

\title{Supplementary Material to ``Evidence for bootstrap percolation dynamics in a photoinduced phase transition"}

\author{Tyler Carbin}
\email{tcarbin@g.ucla.edu}
\affiliation{Department of Physics and Astronomy, UCLA, Los Angeles, CA 90095-1547}

\author{Xinshu Zhang}
\affiliation{Department of Physics and Astronomy, UCLA, Los Angeles, CA 90095-1547}

\author{Adrian B. Culver}
\affiliation{Department of Physics and Astronomy, UCLA, Los Angeles, CA 90095-1547}
\affiliation{Mani L. Bhaumik Institute for Theoretical Physics, Department of Physics and Astronomy, University of California Los Angeles, Los Angeles, CA 90095}

\author{Hengdi Zhao}
\affiliation{Department of Physics, University of Colorado at Boulder, Boulder, Colorado 80309, USA}

\author{Alfred Zong}
\affiliation{Department of Chemistry, University of California at Berkeley, Berkeley, CA, 94720, USA}
\affiliation{Materials Sciences Division, Lawrence Berkeley National Laboratory, Berkeley, CA, 94720, USA}

\author{Rishi Acharya}
\affiliation{Department of Physics and Astronomy, UCLA, Los Angeles, CA 90095-1547}

\author{Cecilia J. Abbamonte}
\affiliation{Department of Physics and Astronomy, UCLA, Los Angeles, CA 90095-1547}

\author{Rahul Roy}
\affiliation{Department of Physics and Astronomy, UCLA, Los Angeles, CA 90095-1547}
\affiliation{Mani L. Bhaumik Institute for Theoretical Physics, Department of Physics and Astronomy, University of California Los Angeles, Los Angeles, CA 90095}

\author{Gang Cao}
\affiliation{Department of Physics, University of Colorado at Boulder, Boulder, Colorado 80309, USA}

\author{Anshul Kogar}
\affiliation{Department of Physics and Astronomy, UCLA, Los Angeles, CA 90095-1547}
\email{anshulkogar@physics.ucla.edu}

\date{\today}
\maketitle

\tableofcontents

\section{Rotational-Anisotropy SHG fitting}

The observed rotational anisotropy (RA) SHG pattern is dominated by the leading-order electric dipole (ED) contribution to the nonlinear susceptibility $\chi^{ED}_{ijk}$, defined by:
\begin{equation}
    P_i(2\omega)=\chi^{ED}_{ijk}E_j(\omega)E_k(\omega),\label{eq:def of chiED}
\end{equation}
where $E_i(\omega)$ is the incident electric field at frequency $\omega$, $P_i(2\omega)$ is the induced polarization in the sample at frequency $2\omega$, and a summation over repeated spatial indices is implied. This contribution to the measured SHG is in general larger than the subleading magnetic-dipole and electric-quadrupole terms by a factor of the incident wavelength divided by the lattice spacing~\cite{FiebigReview}, which here is $\approx$500. 

\begin{figure}
    \centering
    \includegraphics[scale=0.5]{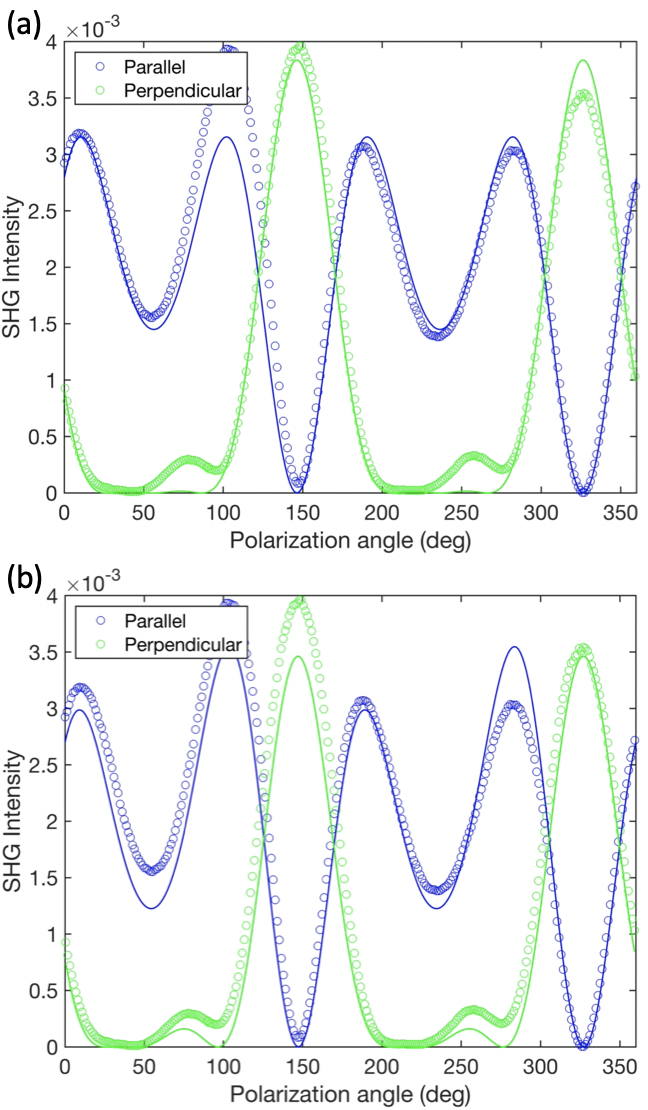}
    \caption{RA-SHG data shown in Fig. 1(a) with \textbf{(a)} a fit to Eq. \eqref{EqRAperfect} (this is the same as the polar plot in Fig. 1(a)) and \textbf{(b)} overlayed with a function given by \eqref{EqRAtilts} with $\theta_a$=$\theta_c$= 0.05 rad. }
    \label{FigRAtilts}
\end{figure}

In general, the nonlinear susceptibility tensor describing ED SHG, $\chi^{ED}_{ijk}\equiv \chi_{ijk}$, is a third-rank tensor with 27 elements. Requiring the tensor to be invariant under all symmetry operations of the crystallographic point group $C_{2v}$ ($mm2$), enforcing the permutation symmetry of indices $j$ and $k$, and taking the axis of inversion symmetry breaking to be the $b$ axis (see Sec. \ref{sec:Symmetry-allowed components of the SHG susceptibility tensor}) leaves 5 independent non-vanishing tensor elements:  

\begin{equation}
\begin{aligned}
    & \chi_{cbc} = \chi_{ccb}, \\
    & \chi_{aba} = \chi_{aab}, \\
    & \chi_{bcc}, \\
    & \chi_{baa}, \\
    & \chi_{bbb}.\label{eq:chi allowed}
\end{aligned}
\end{equation}


The intensity of the second harmonic is proportional to the norm squared of the dot product of $\mathbf{P}(2\omega)$ and the unit vector along the polarization direction of the analyzer $\mathit{\mathbf{e}}(2\omega)$:
\begin{equation}
    I^{2\omega}\propto|\mathit{\mathbf{e}}(2\omega)\cdot \mathit{\mathbf{P}}(2\omega)|^2.
\end{equation}

We then calculate the SHG intensity as a function of the polarization angle $\phi$ of the incident electric field for the cases in which the analyzer is either parallel or perpendicular to the incident light polarization.
In the normal-incidence geometry employed here, the incident electric field is restricted to the $ab$ plane, so that all tensor elements with an index $c$ do not appear in the expression for the SHG intensity. We find:

\begin{equation}
\begin{aligned}
& I^{2\omega}_{Par}(\phi)\propto|\chi_{bbb}\mathrm{cos}^3\phi+(2\chi_{aba}+\chi_{baa})\mathrm{cos}\phi\mathrm{sin}^2\phi|^2 \\
& I^{2\omega}_{Perp}(\phi)\propto|\chi_{baa}\mathrm{cos}^3\phi+(\chi_{bbb}-2\chi_{aba})\mathrm{cos}\phi\mathrm{sin}^2\phi|^2. 
\end{aligned}
\label{EqRAperfect}
\end{equation} 

These equations give the form of the fit shown in Fig. 1(a), where the tensor elements are simultaneously fit in the parallel and perpendicular channels. 
All time traces were measured with polarization angle $\phi$ = 0 in the perpendicular channel such that $I^{2\omega}\propto|\chi_{baa}|^2$.

\subsection{Imperfect alignment}
The measured RA pattern contains small features that can not be captured by Equations \ref{EqRAperfect}. While these features are not relevant to the discussion in the main text, for completeness we show that they can be qualitatively captured by accounting for imperfections in the alignment of the crystal plane to the probe beam. 

The effect of alignment imperfections on the RA-SHG pattern can be approximated by rotating the $\chi^{ED}$ tensor by small angles before calculating $I^{2\omega}(\phi)$. 
This significantly complicates the functional form of $I^{2\omega}$ and introduces terms containing elements $\chi_{ccb}$ and $\chi_{bcc}$ that are absent when the probe beam is exactly perpendicular to the $ab$ plane of the crystal.

Figure \ref{FigRAtilts}(b) shows the RA data overlayed with a functional form given by Eq. \ref{EqRAtilts} below. These functional forms account for tilts of the crystal about the $a$ and $c$ axes away from the ideal orientation in which the $ab$ plane is exactly normal to the probe beam. Importantly, we find that the bumps in the perpendicular pattern at 80$\deg$ and 260$\deg$ can be qualitatively accounted for. 

\begin{widetext}

\begin{equation}
    \begin{aligned}
I^{2\omega}_{Par}(\phi)\propto&|(\mathrm{cos}\theta_c\mathrm{cos}\theta_a\mathrm{cos}\phi-\mathrm{sin}\theta_c\mathrm{sin}\phi)
(\mathrm{cos}^2\phi((2\chi_{aba}+\chi_{baa})\mathrm{cos}^2\theta_a\mathrm{sin}^2\theta_c+
(2\chi_{cbc}+\chi_{bcc})\mathrm{sin}^2\theta_a)+\\
&(2\chi_{aba}+\chi_{baa}-\chi_{bbb})\mathrm{cos}\theta_a\mathrm{sin}(2\theta_c)\mathrm{cos}\phi\mathrm{sin}\phi+
\chi_{bbb}\mathrm{sin}^2\theta_c+\mathrm{cos}^2\theta_c(\chi_{bbb}\mathrm{cos}^2\theta_a\mathrm{cos}^2\phi+
(2\chi_{aba}+\chi_{baa})\mathrm{sin}^2\phi))|^2 \\
I^{2\omega}_{Perp}(\phi)\propto&|\frac{1}{2}(-\mathrm{cos}\theta_c(-2\chi_{aba}-\chi_{baa}-\chi_{bbb}+
(2\chi_{aba}+\chi_{baa}-\chi_{bbb})\mathrm{cos}(2\theta_c))\mathrm{cos}^3\theta_a\mathrm{cos}^2\phi\mathrm{sin}\phi+\\
&\mathrm{cos}^2\theta_a\mathrm{sin}\theta_c\mathrm{cos}\phi((2\chi_{aba}-\chi_{baa}-\chi_{bbb}+
(2\chi_{aba}+\chi_{baa}-\chi_{bbb})\mathrm{cos}(2\theta_c))\mathrm{cos}^2\phi-\\
&2(\chi_{baa}-\chi_{bbb}+(2\chi_{aba}+\chi_{baa}-\chi_{bbb})\mathrm{cos}(2\theta_c))\mathrm{sin}^2\phi)+\mathrm{cos}\theta_c\mathrm{cos}\theta_a\mathrm{sin}\phi
((-2\chi_{cbc}-\chi_{bcc}-2\chi_{baa}+2\chi_{bbb}+\\
&2(2\chi_{aba}+\chi_{baa}-\chi_{bbb})\mathrm{cos}(2\theta_c)+
(2\chi_{cbc}+\chi_{bcc})\mathrm{cos}(2\theta_a))\mathrm{cos}^2\phi-\\
&(-2\chi_{aba}+\chi_{baa}+\chi_{bbb}+(2\chi_{aba}+\chi_{baa}-\chi_{bbb})\mathrm{cos}(2\theta_c))\mathrm{sin}^2\phi)+\mathrm{cos}\phi\mathrm{\theta_y}(-2\chi_{bcc}\mathrm{cos}^2\phi\mathrm{sin}^2\theta_a-\\
&(-2\chi_{cbc}+2\chi_{aba}+\chi_{baa}+\chi_{bbb}+(2\chi_{aba}+\chi_{baa}-\chi_{bbb})\mathrm{cos}(2\theta_c)+
2\chi_{cbc}\mathrm{cos}(2\theta_a))\mathrm{sin}^2\phi))|^2. 
\end{aligned}
\label{EqRAtilts}
\end{equation}
\end{widetext}

\subsection{Multiple Domains}
\label{SsecMultiDom}

If there are multiple domains in the probed region of the sample, the total SHG intensity at polarization angle $\phi$ is given by
\begin{equation}
    I^{2\omega}_{tot}(\phi) = \sum_i f_i I^{2\omega}_i(\phi),
\end{equation}
where the sum is over unique domains and $f_i$ gives the fraction of the probed region of the sample that constitutes the $i^{\mathrm{th}}$ domain. 

By referring to Eq. \eqref{EqRAperfect}, we see that if there exist two domains related by $\phi\rightarrow\phi+$90\degree, then there exists no angle for which the total SHG intensity $I^{2\omega}_{tot}(\phi)$ is zero. Domains related by a 90$\degree$ rotation of the polar moment therefore lift the nodes in the RA-SHG pattern and are not present in the data presented here.  

On the other hand, the 180$\degree$ polar domains can only change the magnitude of $I^{2\omega}$ as a function of position in the $ab$ plane. This is discussed at length in Ref.~\cite{Lei}. Thus, one cannot reliably compare the magnitude of the pump-induced drop in $I^{2\omega}$ between data-sets (it varied between $\sim$4-40\% in our measurements, but was typically $\sim$15\%). Nonetheless, at a single spot, the jump in $I^{2\omega}$ is constant, which indicates that the 180$\degree$ domains are static upon photo-excitation and that the dominant dynamics are associated with the insulator-metal transition. 


\section{SHG Wavelength Dependence} \label{SsecWavelength}

\begin{figure}[h]
    \centering
    \includegraphics[scale=0.6]{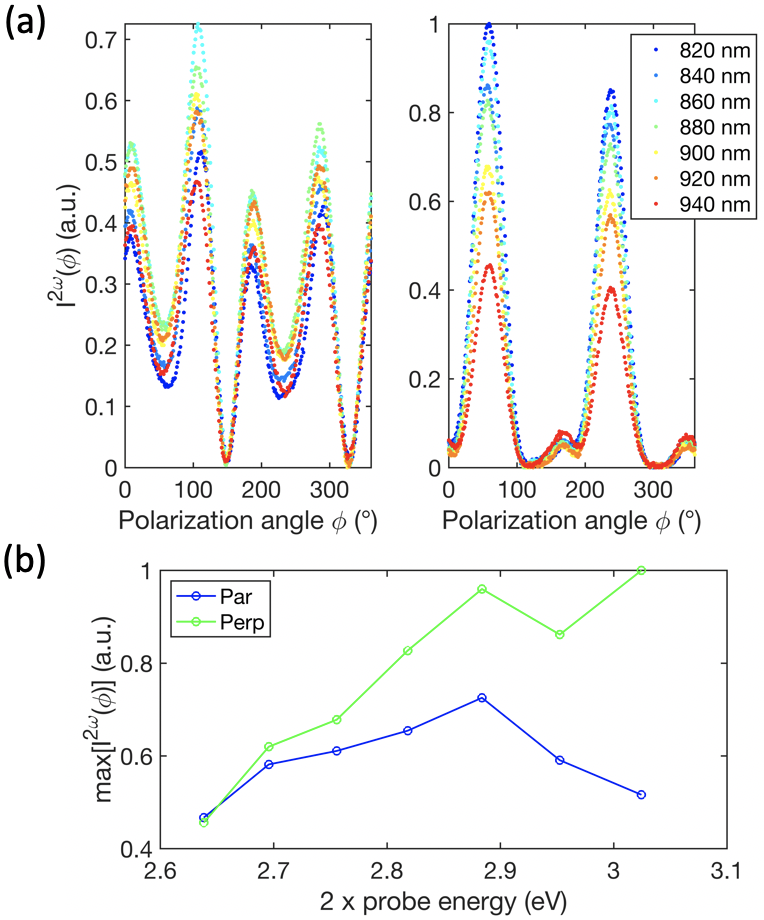}
    \caption{\textbf{(a)} Wavelength dependence of the RA-SHG measurements in both parallel (left) and perpendicular (right) polarization channels. There is a 90$\degree$ offset between the two channels. \textbf{(b)} The maximal value in the $I^{2\omega}$ curves from \textbf{a} plotted vs. twice the probe energy.}
    \label{fig:wavelengthScan}
\end{figure}

Figure \ref{fig:wavelengthScan}(a) shows the dependence of the RA-SHG pattern on the wavelength of the probe beam between 820nm and 940nm. Fig. \ref{fig:wavelengthScan}(b) shows the value of $\textrm{max}(I^{2\omega}(\phi))$ in the two channels. The only change in the pattern comes from the dependence of the tensor elements on the light wavelength. 
The same fitting function Eq.~\ref{EqRAtilts} can be used to fit the RA-SHG pattern at each wavelength.

\section{Time Trace Fits} \label{SsecFitTT}

All time traces are fit to Eq.~1 of the main text. For the data point at the largest fluence (1.2 mJ/cm$^2$), $\tau$ was restricted to be larger than 100 fs so that the fitting procedure would converge to a solution. Because the best-fit value of $\tau$ is found to be $\approx$170 fs (i.e. larger than this lower bound), we do not expect that this constraint significantly altered the extracted value. However, it is possible that this constraint resulted in an underestimation of the uncertainties of the extracted $\tau$ and $\alpha$ at this fluence. 

\begin{figure*}[htb!]
    \centering
    \includegraphics[scale=0.60]{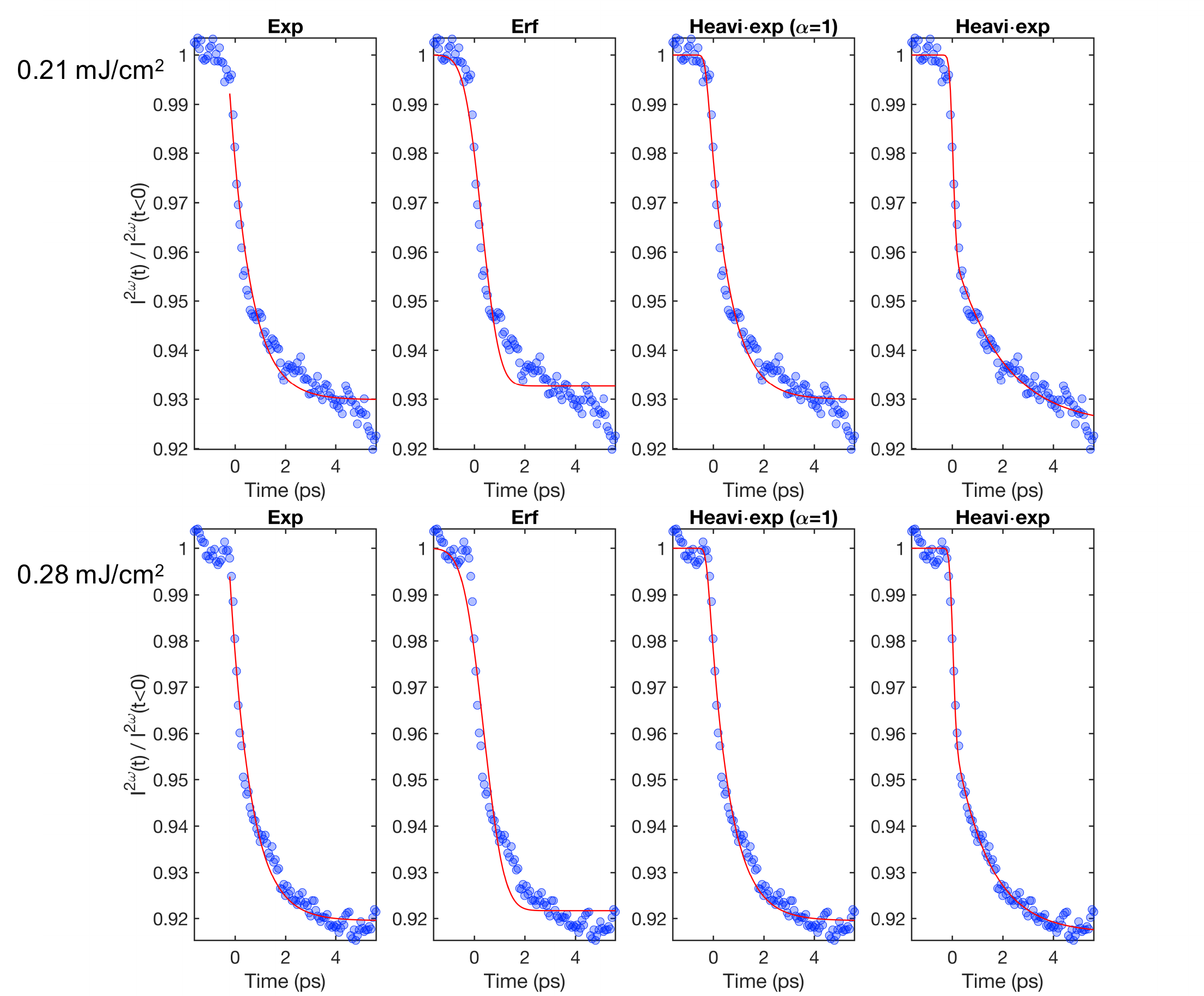}
    \caption{Attempted fits to the 0.21 mJ/cm$^2$ (top) and 0.28 mJ/cm$^2$ (bottom) time traces shown in Fig. 3 using various fitting functions. The left column shows the attempted fit of all time points after $t_0$ to a simple exponential decay with variable decay constant. The second column shows attempted fits to an error function of variable width with a convolution of the pump-probe cross correlation. The third column shows the fits to Eq. 1 with $\alpha$ fixed to one. The right column shows the fits to Eq. 1 that are used in the main text.}
    \label{fig:badFits}
\end{figure*}

There are two timescales in Eq. 1 that correspond to physical processes in the sample: $\tau$ and a shorter timescale represented by the Heaviside function that is smaller than, or comparable to our experimental time resolution $\approx$~280~fs (c.f. Section~\ref{Ssec:timeRes}). 
Fig. \ref{fig:badFits} shows that the inclusion of only one of these timescales without the other is insufficient to obtain a good fit to the measured time traces and thus that the presence of both in Eq. 1 is necessitated by the data.
The left panel of Fig. \ref{fig:badFits} shows that fits lacking a short timescale to govern the immediate drop in intensity after the pump pulse cannot capture the data well. In particular, the fits shown are of the data after $t_0$ to a simple exponential decay with no convolution with the cross-correlation of the pump and probe pulses $g(w_0,t)$.
Similarly, to demonstrate that the time constant $\tau$ is necessary to obtain a good fit, we present attempted fits to Eq. \ref{FErfNoExp} in the second column of Fig. \ref{fig:badFits}. This function has no exponential decay but includes an error function of variable width $w$ that approximates the Heaviside function in the limit that $w\rightarrow$0. 
Lastly, the third column displays attempted fits to Eq. 1 with $\alpha$ fixed to one. This demonstrates that, to obtain a satisfactory fit, the exponential term must have a different weight from the term that has only the Heaviside function. 

\begin{multline}
     u(t)=1+\\
     \Biggl[\frac{1}{2}\left(1+\mathrm{Erf}\left(\frac{2\sqrt{\mathrm{ln}(2)}(t-t_0)}{w}\right)\right)I_{\infty}\Biggr] \ast g(w_0,t).
     \label{FErfNoExp}
\end{multline}

\section{Recovery time traces}
\begin{figure}[h]
    \centering
    \includegraphics[scale=0.5]{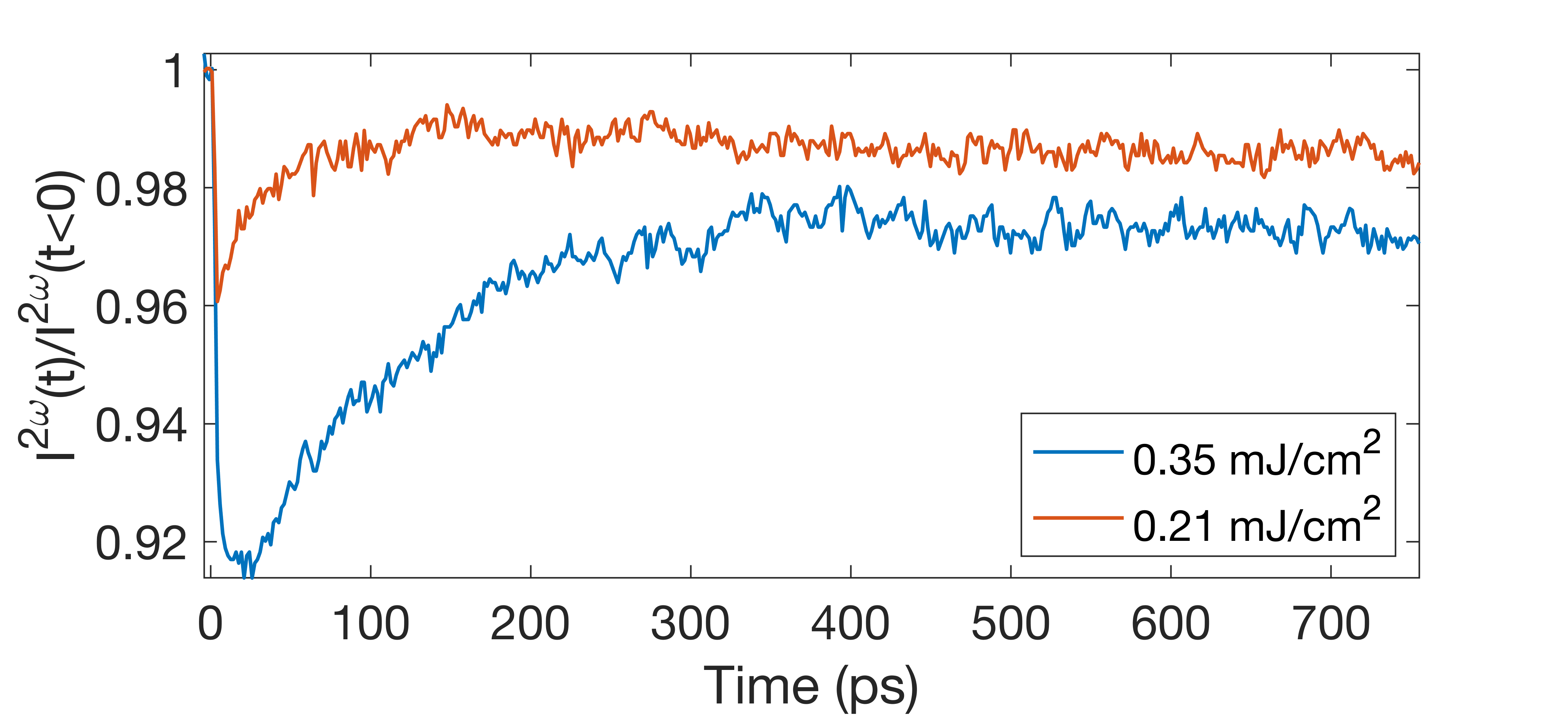}
    \caption{SHG time traces taken over 100s of ps to show the recovery between pulses. 
    }
    \label{fig:recovery}
\end{figure}

Figure \ref{fig:recovery} shows the recovery of $I^{2\omega}$ following the pump pulse. These traces show that the recovery timescale (hundreds of ps) is much longer than the timescale for the initial insulator-metal transition, which justifies our separation of timescales in the percolation model.




\section{Estimates of Time Resolution}
\label{Ssec:timeRes}
In addition to the intrinsic widths of the pump and probe pulses, the time resolution of the experiment is affected by the angle mismatch between the two beams.
In the limit in which both the pump and probe beams are normally incident on the sample, the cross-correlation between the two Gaussian profiles each with a $\sigma$=77~fs (181~fs FWHM) is another Gaussian with $\sigma$=$\sqrt{2}(77$~fs) = 109~fs (256~fs FWHM).
For the geometry employed in the experiment, we can obtain an upper bound on the time resolution by adding the time interval between the arrival of the pump pulse at the near and far ends of the transverse probe profile to the ``intrinsic" width of the pump pulse. 
This correction gives: $\Delta t$ = FWHM$_{probe}$sin(15$\degree$)/c = 35~fs.
Thus, the obtained upper bound for the FWHM of the cross-correlation between the pump and probe pulses in the experiment is $\sqrt{(181)^2+(181+35)^2}$~fs = 282~fs.

\begin{figure}
    \centering
    \includegraphics[scale=0.4]{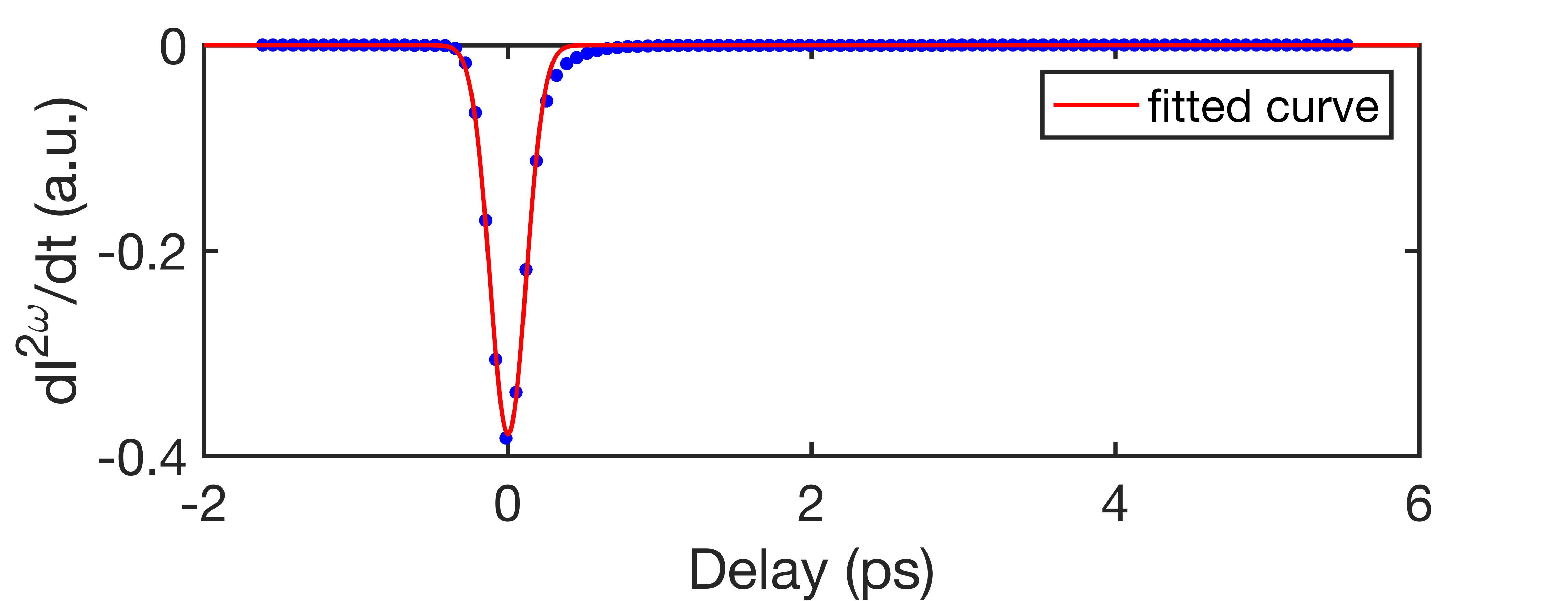}
    \caption{The time-derivative of the fitted curve shown in Fig~3 in the main text for 1.2~mJ/cm$^2$ (blue points) and a gaussian fit (red) used to determine the experimental time resolution.}
    \label{figTimeResDeriv}
\end{figure}

We can also estimate the experimental time resolution from the fitted time traces and compare it to this upper bound. 
To do this, we fit the derivative of the time trace corresponding to the largest fluence (1.2~mJ/cm$^2$) to a gaussian. 
If we assume that the drop in SHG intensity occurs much faster than the time resolution of our experiment, the width of the fitted gaussian is a measure of our experimental time resolution. (Otherwise, it is an upper bound). 
The derivative of the fitted curve for the time trace corresponding to the largest fluence and the gaussian fit to this derivative are shown in Fig.~\ref{figTimeResDeriv}.
The experimental time resolution determined from this procedure is \textbf{278~fs} (given by the FWHM of the gaussian). 
This value is in remarkable agreement with the upper bound estimated above, 282~fs.  

\section{Calculation of $d_\gamma$}
\label{Ssec:dgamma}



We use the following equations to calculate the number of photons absorbed by a single pulse and the number of unit cells impinged by a single pulse
\begin{equation}
    N_\gamma = (1-R)F\frac{1}{h\nu}\pi(\textrm{hwhm}_{pump})^2,
    \label{Ngamma}
\end{equation}
\begin{equation}
    N_{u.c.} = \frac{\pi(\textrm{hwhm}_{pump})^2\delta(\omega_{pump})}{abc},
    \label{Nuc}
\end{equation}
where $R$ is the reflectivity of CRO at 1030 nm, $h\nu$ is the energy of a single 1030~nm photon, and $\delta_{pump}\approx$ 78~nm is the penetration depth of the pump laser (c.f. Section~\ref{SSecPenDepth}).
We find the number of unit cells per absorbed pump photon to be $N_{u.c.}/N_{\gamma}$ = 11.5 at the largest $\tau$. 

We can then multiply by the average lattice parameter $\overline{a}\equiv \frac{a+b+c}{3}$ = 10.1 $\mathrm{\AA}$ to obtain the average distance between absorbed pump photons $d_\gamma$:

\begin{equation}
    d_{\gamma}\approx \frac{N_{u.c.}}{N_{\gamma}}\overline{a}.
    \label{eq:dgamma}
\end{equation}





\section{Penetration depth calculations}
\label{SSecPenDepth} 
To determine the penetration depths of the pump laser (1030~nm) and of the first- (900~nm) and second- (450~nm) harmonic probe beams, we performed reflectivity measurements of the sample at each of these wavelengths. 
To verify this method, we also measure the reflectivity at 800~nm for comparison with the penetration depth reported in Ref.~\cite{Lei}.
We use these reflectivity measurements in conjunction with previous measurements~\cite{Lei} of the real part of the conductivity, $\sigma_1$, to determine the penetration depths. 

The reflectivity $R$ is related to the optical conductivity by the following expression (in cgs units):
\begin{equation}
    R=\frac{1+\frac{4\pi}{\omega}(\sigma_1^2+\sigma_2^2)^{1/2} - (\frac{8\pi}{\omega})^{1/2}\left((\sigma_1^2+\sigma_2^2)^{1/2}+\sigma_2\right)^{1/2}}{1+\frac{4\pi}{\omega}(\sigma_1^2+\sigma_2^2)^{1/2} + (\frac{8\pi}{\omega})^{1/2}\left((\sigma_1^2+\sigma_2^2)^{1/2}+\sigma_2\right)^{1/2}},
    \label{eq:reflectivity}
\end{equation}
where $\sigma_{1(2)}$ is the real(imaginary) part of the optical conductivity. 
From this, we extract $\sigma_2(\omega)$ at the desired frequencies, which we use to compute the penetration depth $\delta$:
\begin{equation}
    \delta = \frac{c}{(2\pi\omega)^{1/2}}\left((\sigma_1^2+\sigma_2^2)^{1/2}+\sigma_2\right)^{-1/2},
    \label{eq:penDepth}
\end{equation}
where we have taken $\mu_1$=1. 

The measured values of the reflectivity $R_{meas}$ are included in Table~\ref{table:penDepths}. 
Of the relevant wavelengths, only the reflectivity at 1030~nm has been previously measured~\cite{Lee07}, and the value reported here (0.18) is in good agreement with these prior results.
However, for all included wavelengths, the measured reflectivity is inconsistent with the previously reported values for $\sigma_1$. In particular, using the values for $\sigma_1$ in the literature and the measured reflectivity, Eq.~\ref{eq:reflectivity} yields a complex value for $\sigma_2$, which is defined to be real. This result is unphysical. 
Substituting larger values for the reflectivity, however, yields real values for $\sigma_2$ at all considered wavelengths.
To determine the penetration depth, we use the minimum reflectivity ($R_{det}$) that results in a real-valued (and thus, physical) $\sigma_2$. (In all cases, the difference with the measured reflectivity was less than or equal to 3$\%$).
This procedure is validated by the penetration depth calculated using this method at 800~nm (73~nm), which is in exact agreement with that reported in Ref.~\cite{Lei}. The calculated penetration depth at 450~nm is also similar to that reported at 400~nm~\cite{Lei}.
The results of the penetration depth calculations are summarized in Table~\ref{table:penDepths}.

\begin{table}[h]
\centering
    \begin{tabular}{|c||c|c|c||c|}
    \hline
    $\lambda$ (nm)  & $\sigma_1$ (THz) & R$_{det}$(R$_{meas}$) & $\sigma_2$ (THz)& $\delta$ (nm)\\
      \hline
      800  & 681 & 0.19(0.16) & 373 & 73\\
      \hline
      450 & 1330 & 0.20 & 930 & 31\\ 
      \hline 
      900 & 614 & 0.19(0.16) & 365 & 76\\
      \hline 
      1030 & 576 & 0.20(0.18) & 576 & 78\\
      \hline
    \end{tabular}
    \caption{The penetration depth $\delta$ in CRO at the pump wavelength 1030~nm, at the first- (900~nm) and second- (450~nm) harmonic probe wavelengths, and at 800~nm for comparison to Ref.~\cite{Lei}. The real and imaginary parts of the optical conductivity, and the measured and determined values for the reflectivity that are used to calculate the penetration depth are also included. }
    \label{table:penDepths}
\end{table}

\section{Considerations of homogeneous mechanisms}
\label{SSecAltMech}

In the main text we have argued for the existence of percolating inhomogeneity in the PIPT by correlating our measurements with simulations of inhomogeneous dynamics. 
While the possibility of a homogeneous PIPT mechanism cannot be excluded by these results, we argue here that such a scenario is unlikely. 
The most essential feature of our data that appears to require the invocation of spatial inhomogeneity is the non-monotonic fluence dependence of the time constant $\tau$. If the PIPT is thought to be homogeneous, additional description is needed to explain this behavior.

One recent study~\cite{ZongDSD} used a time-dependent Ginzburg-Landau formalism to describe the non-monotonic fluence dependence of the photo-induced transition time in a charge density wave compound. In this picture, the slowdown in $\tau$ appears due to fluctuations with correlation lengths that diverge near a critical fluence. 
One might suspect that a similar description applies here. However, the PIPT observed in CRO appears substantially different from the previous study in several respects. 

First, in the previous study there is only one observed timescale governing the photoinduced transition due to the coherent displacive motion of the atoms. It is this timescale that exhibits the slowing down behavior. 
In comparison, there exist at least two timescales involved in the transition process measured in this work (Fig. \ref{fig:badFits}). 
While the presence of two timescales is an essential feature of our inhomogeneous model (being associated with the quasi-instantaneous structural change associated with the absorption of light and with the subsequent percolation), there is no apparent analogue to the longer timescale in the homogeneous Ginzburg-Landau picture. 
Additionally, the change in the decay time constant observed in the previous work is much smaller ($\sim$300-400 fs) than that seen here, despite both transitions involving changes to the crystal lattice. This observation suggests that the coherent motion of atoms cannot explain the exponential part of the dynamics, which is consistent with mesoscale structural inhomogeneity in CRO.
In summary, unlike a Ginzburg-Landau description, mesoscopic percolation in CRO offers a natural explanation for (1) why there exist two timescales in CRO and (2) why the timescale of the decay in CRO should be significantly longer than that in the previous study~\cite{ZongDSD}. 

We have addressed the possibility of an explanation based on a time-dependent Ginzburg-Landau theory in detail because it is, to our knowledge, the most straightforward way in which one might attempt to attribute our observations to homogeneous dynamics. There are surely other possibilities that we have not considered. 
However, we emphasize several reasons why the appearance of mesoscale inhomogeneity may be expected: (1) the well-documented appearance of inhomogeneity in the vicinity of first-order insulator-metal transitions, (2) the central role inhomogeneity plays in the observed colossal magnetoresistance effect, which is observed in CRO, (3) the observation of inhomogeneity in the Ti-doped compound~\cite{McLeod21}, and (4) that the nearly binary nature of the $c$ axis change observed in equilibrium is likely to persist away from equilibrium~\cite{Perez-Salinas2022}.




\section{Simulation length and time scales}
\label{sec:simScales}

Although the simulations alone are devoid of any length/time scale, we can extract crude estimates for the physical scales that correspond to elements of the simulation by comparing the simulation output with the data. 
By equating the peak value of $\tau$ from the measurements with that from the simulation, we get that one time step in the simulation corresponds to several hundred fs. 
Using this value, and equating the percolation speed to the ``percolation speed" of the simulations (1 site per time step), we find that the length scale corresponding to a single site in the simulation is on the order of 1~nm. 

\section{Calculation of ``strain" in the simulations}
The ``strain" in the simulations (that dictates a switch to the $L_c$ state when greater than or equal to $\sigma_{th}$) is calculated according to the following equations: 

\begin{equation}
\begin{aligned}    
    \sigma_{3D}(i,j,k)= &s_{i+1,j,k}+s_{i,j+1,k}+s_{i,j,k+1}+\dots \\
    &s_{i-1,j,k}+s_{i,j-1,k}+s_{i,j,k-1} +\dots \\
    \beta(&s_{i+1,j+1,k}+s_{i-1,j+1,k}+s_{i+1,j-1,k}+\dots \\
    &s_{i-1,j-1,k}+s_{i+1,j,k+1}+s_{i-1,j,k+1}+\dots \\ &s_{i+1,j,k-1}+s_{i-1,j,k-1}+s_{i,j+1,k+1}+\dots \\ &s_{i,j-1,k+1}+s_{i,j+1,k-1}+s_{i,j-1,k-1}), \\
    \sigma_{2D}(i,j)=&s_{i,j+1}+s_{i,j-1}+s_{i+1,j}+s_{i-1,j} +\dots \\
    \beta(s_{i+1,j+1} & +s_{i-1,j+1}+s_{i-1,j-1}+s_{i-1,j-1}).
    \label{simStrain}
\end{aligned}
\end{equation}
where $\sigma_{2D}$($\sigma_{3D}$) is used for a two-(three-)dimensional simulated sample. $s_{i,j,k}\in \{0,1\}$ is the state of site $(i,j,k)$ where 0 and 1 correspond to the $L_c$ and $S_c$ states, respectively. These equations also define the next-nearest neighbor coupling with a coupling strength given by the parameter $\beta\in$[0,1] (which is set to zero in the simulations included in the main text). 

\section{Fits to simulated time traces}

\begin{figure}[h!]
    \centering
    \includegraphics[scale=0.3]{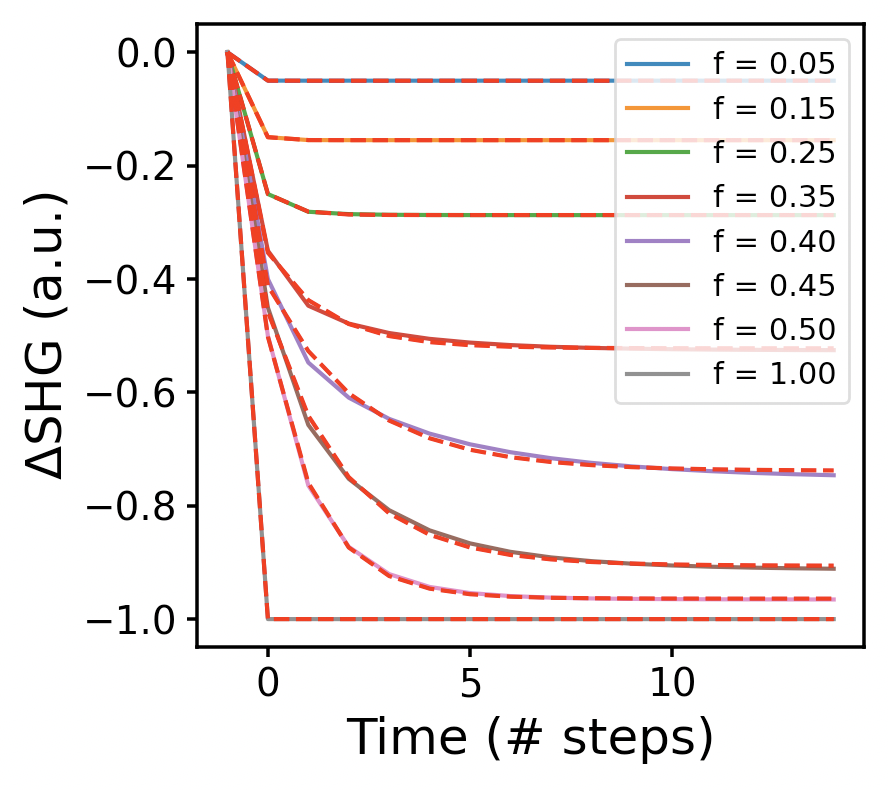}
    \caption{Fits (dashed red curves) of simulated time traces shown in the main text to Eq.~1 without a convolution with the cross correlation factor.}
    \label{SimTTFits}
\end{figure}

The fits to the simulated time traces were performed with a procedure almost identical to the fitting of the experimental data. The only difference is the removal of the cross correlation factor, which accounts for the temporal resolution associated with the laser pulses. Fig.~\ref{SimTTFits} shows fits of the simulated time traces to Eq.~1 of the main text excluding the cross-correlation term.

\section{\label{SecRobustSim}Robustness of Simulation results}

Bootstrap percolation models have been widely studied in the mathematics literature. In the thermodynamic limit, the entire system becomes activated at a fraction of initialized sites $p_c$ which approaches 0 or 1 depending on the dimension and the number of nearest neighbors required for activation~\cite{Schonmann92}. However, finite-size effects are found to dominate the dynamics of even the largest computer simulations. Indeed the threshold of percolation has been found to approach the thermodynamic limit as slowly as $\frac{C}{\mathrm{log}\left(\mathrm{log}(n)\right)}$ for a system of size $n^3$ in some of the models that we have studied here~\cite{balogh_sharp_2012,balogh_bootstrap_2009}.
Thus, the percolation process undergone during the PIPT studied here is almost certainly far from the thermodynamic limit, and is best studied via computer simulation. 

In this section we show that variations of the simulation discussed in the main text produce time traces that preserve the salient qualitative features of the traces. In particular, the peak in the time constant $\tau$ as a function of fluence fraction $f$ and its alignment with the saturation of $I_{\infty}$, as well as the $f$-dependence of $\alpha I_\infty$ that is non-monotonic and goes to zero at extremal values of $f$, are reproduced in the various simulations shown below. These features should therefore be understood as rather general consequences of our proposed model as opposed to artifacts of the specific settings with which the simulation in the main text was performed. 

\subsection{Varying $\sigma_{th}$}

\begin{figure}
    \centering
    \includegraphics[scale=0.42]{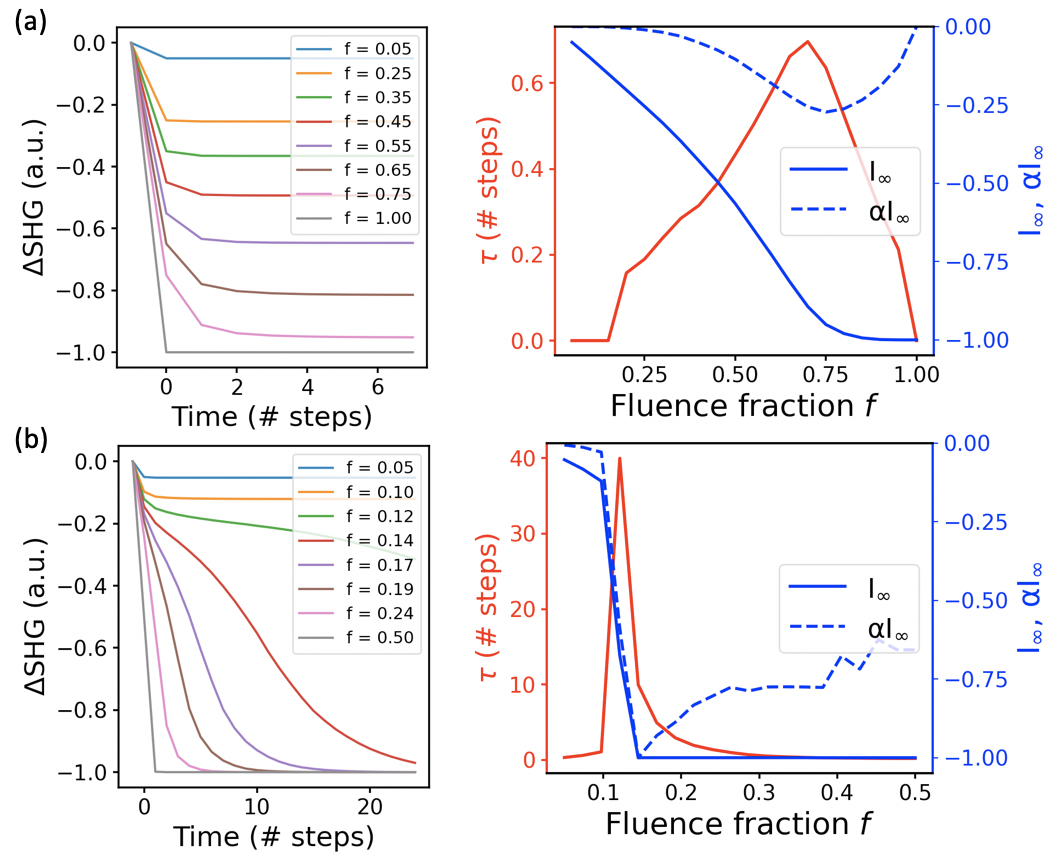}
    \caption{Simulated time traces and extracted fit parameters with $\sigma_{th}$ = 4 (top) and $\sigma_{th}$ = 2 (bottom) (with $\beta$ = 0). Note that the $x$-axes are not the same in the two cases.}
    \label{varySth}
\end{figure}

Figure \ref{varySth} shows the output of simulations using a 30x30x30 array with $\sigma_{th}$ changed from 3 as in the main text to 2 and to 4.
(No significant differences were observed between simulations using a 30x30x30 and a 40x40x40 array.) 
As may be expected, the value of $f$ at which $\tau$ peaks and $I_{\infty}$ saturates increases with $\sigma_{th}$.

While the $\sigma_{th}$ = 4 case retains time traces that appear similar to those in the data and in the $\sigma_{th}$ = 3 case, the $\sigma_{th}$ = 2 case appears qualitatively different.
For the $\sigma_{th}$=~2 case and more generally for $\sigma_{th} <d$ where $d$ is the dimension of the model, one expects a sharp transition for large enough systems based on rigorous results in the mathematics literature~\cite{holroyd_sharp_2003,balogh_sharp_2012}. Our simulations for higher values of $\sigma_{th}$ seem to indicate the lack of a sharp transition at least at the system sizes that we have simulated. 
We take this as evidence that the experiment is best modeled with $\sigma_{th}$ = $d$.
Nonetheless, in all cases the general behavior of the three fit parameters is preserved.

\subsection{2$d$ Simulations}
\begin{figure}[h]
    \centering
    \includegraphics[scale=0.42]{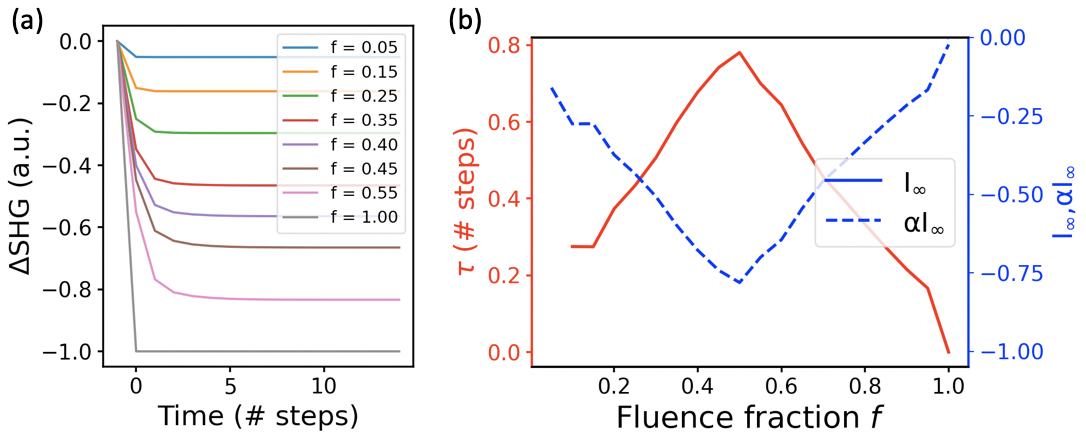}
    \caption{Simulated SHG time traces \textbf{(a)} and extracted fit parameters \textbf{(b)} for a 2d 30x30 simulation with $\sigma_{th}$=2 and $\beta$=0.}
    \label{2dSimbeta0}
\end{figure}

The broad qualitative features of the simulated time traces are unchanged when the simulation is limited to two spatial dimensions. The results of the 2d simulation for $\sigma_{th}$=2 and $\beta$=0 are shown in Fig. \ref{2dSimbeta0}.

Fig. \ref{2dSimbeta0} establishes that our model of percolating inhomogeneity reproduces the qualitative features observed in the experiment even if the simulation is confined to two dimensions.

\subsection{Next-nearest neighbors}

\begin{figure}
    \centering
    \includegraphics[scale=0.42]{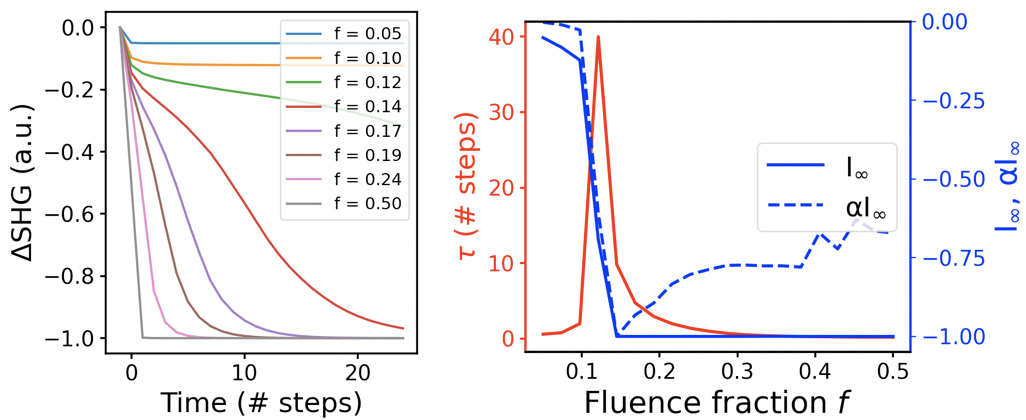}
    \caption{Simulated time traces and extracted fit parameters for a 30x30x30 array with $\sigma_{th}$ = 3 and $\beta$ = 0.1.}
    \label{Sth3betap1}
\end{figure}

We introduce coupling between next-nearest neighbors by setting $\beta>$ 0 in Eq.~\ref{simStrain}. Figure \ref{Sth3betap1} shows the simulated time traces and extracted parameters for the 30x30x30 simulation with $\sigma_{th}$ = 3 and $\beta$ = 0.1. 
Again we find that, while the simulated time traces appear more complicated than the data, the $f$-dependence of the parameters is qualitatively unchanged. 

\begin{figure}
    \centering
    \includegraphics[scale=0.42]{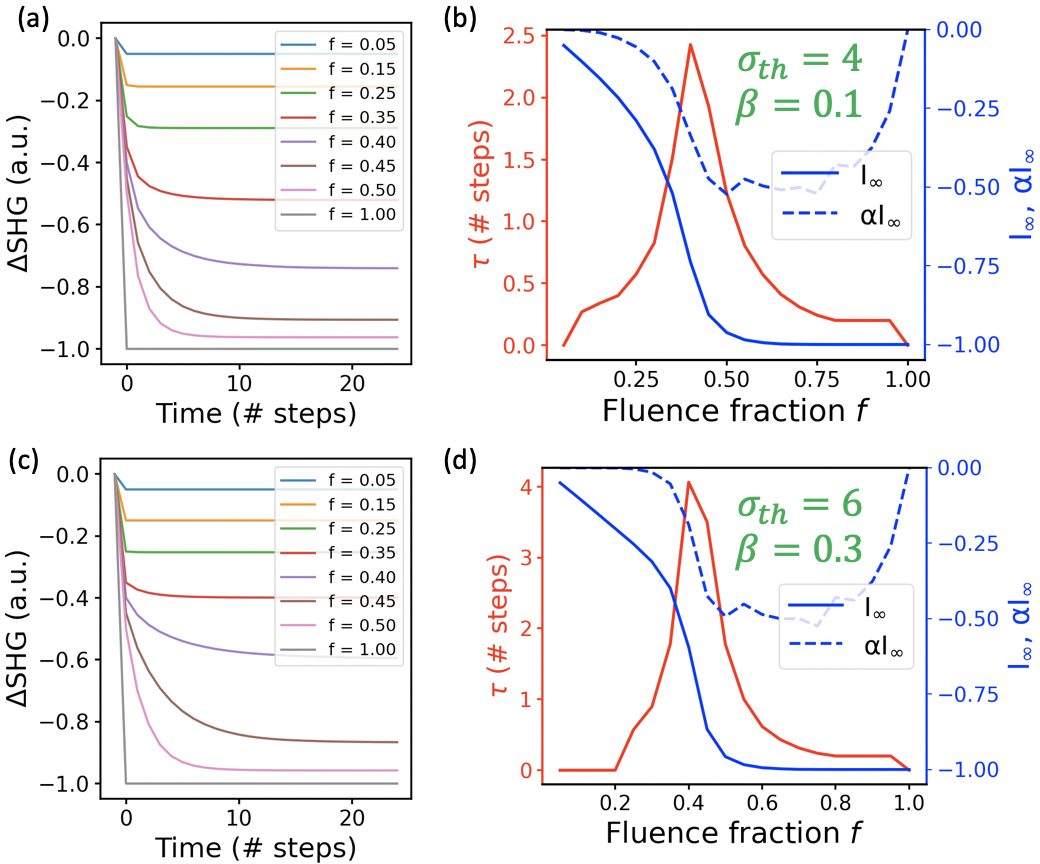}
    \caption{Simulated time traces and extracted fit parameters for $3d$ 30x30x30 array simulations with \textbf{(a)(b)} $(\sigma_{th}, \beta)$=(4,0.1) and \textbf{(c)(d)} $(\sigma_{th}, \beta)$=(6,0.3) }
    \label{ExtraSims3d}
\end{figure}

\begin{figure}
    \centering
    \includegraphics[scale=0.42]{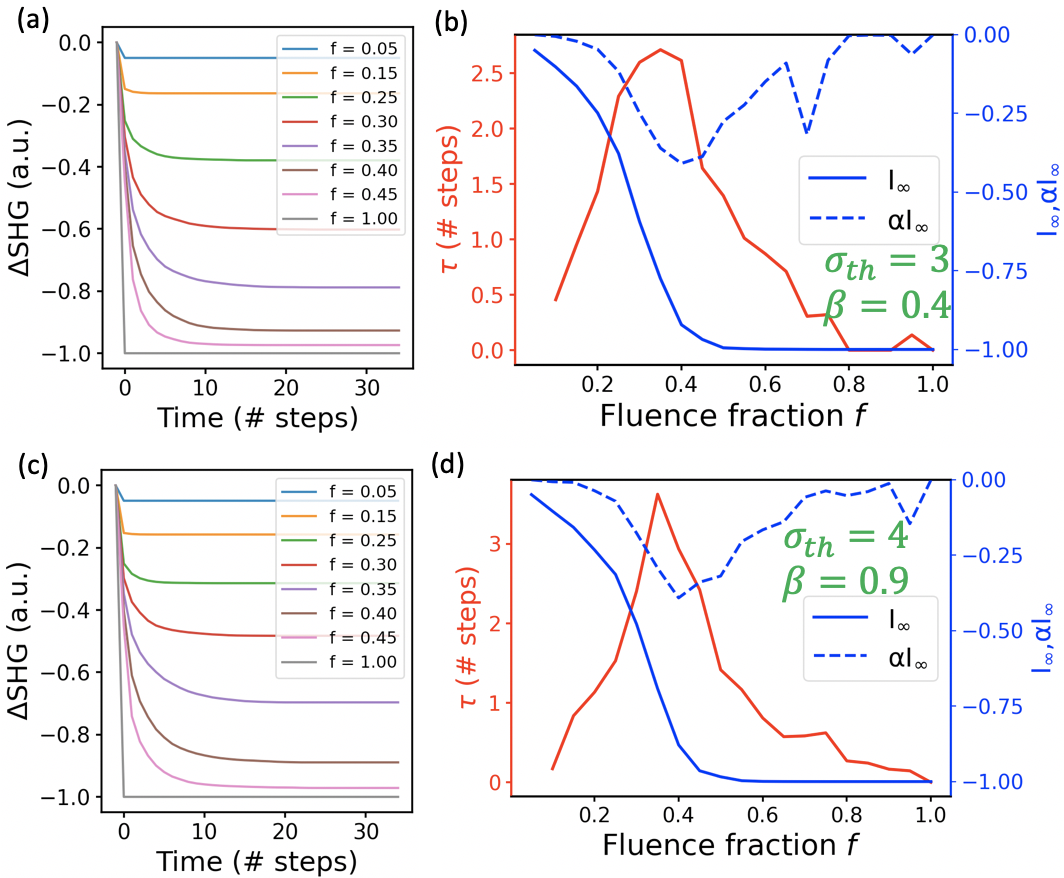}
    \caption{Simulated time traces and extracted fit parameters for $2d$ 30x30 array simulations with \textbf{(a)(b)} $(\sigma_{th}, \beta)$=(3,0.4) and \textbf{(c)(d)} $(\sigma_{th}, \beta)$=(4,0.9) }
    \label{ExtraSims2d}
\end{figure}

One may be tempted to interpret this as indicative that the next-nearest neighbor coupling between sites is insignificant in the PIPT in CRO. 
However, by increasing both $\sigma_{th}$ and $\beta$, one finds that there are many combinations of the two parameters (in both $2d$ and $3d$) that appear to reproduce the qualitative features of the data. The results of several of these simulations are shown in Figures \ref{ExtraSims3d} and \ref{ExtraSims2d} for the $3d$ and $2d$ cases respectively. 

\subsection{Penetration depth effects}

In principle, the finite penetration depths of the pump and probe lasers introduce another type of inhomogeneity into the PIPT. Namely, the number of (pump and probe) photons absorbed by the sample falls off exponentially as a function of the depth from the surface. In this section we demonstrate that the qualitative results of the simulations are preserved when this additional inhomogeneity is accounted for. 

The intensity of the pump beam as a function of depth $z$ falls off exponentially with the penetration depth $\delta(\omega_{pump})$:
\begin{equation}
    I_{pump}(z)\propto e^{-z/\delta(\omega_{pump})}.
\end{equation}

In the simulations, this effect is captured by scaling the probability of activating a site at $t=0$ (given by the fluence fraction $f$) by this exponential factor, where $z$ is now the discrete $z$-coordinate in the simulated array, and the penetration depth is an unknown number of array sites (discussed further below). 

To account for the finite penetration depth of our SHG probe, we must consider both the penetration depth of the incident laser at frequency $\omega$ and that of the generated second harmonic 2$\omega$.
The second-harmonic field generated at a depth $z$ is proportional to the square of the field at position $z$ with frequency $\omega$, which decays exponentially with the penetration depth $\delta(\omega)$:
\begin{equation}
    E^{2\omega}(z)\propto E^\omega(z)^2 \propto e^{-z/\delta(\omega)}.
\end{equation} 

The generated second harmonic intensity is then partially absorbed by the sample before returning to the surface.
Thus, the light that contributes to the measured SHG intensity is additionally damped along the $z$ direction due to the penetration depth of the second harmonic. 
This leads to an effective penetration depth of the SHG probe:
\begin{equation}
    I^{2\omega}_{contr}(z)\propto e^{-z/\delta(2\omega)}e^{-z/\delta(\omega)}= e^{-z/\delta^{eff}},
\end{equation}
where we have defined the effective penetration depth:
\begin{equation}
    \delta^{eff}\equiv\frac{\delta(\omega)\delta(2\omega)}{\delta(\omega)+\delta(2\omega)}.
    \label{eq:effPenDepth}
\end{equation}
 
To incorporate the effect of this finite-size penetration depth into the simulations, the simulated SHG intensity, which was previously simply proportional to the number of $S_c$ sites, is now proportional to the number of $S_c$ sites weighted by $e^{-z/\delta^{eff}}$, where $z$ is the discrete z-coordinate in the array and $\delta^{eff}$ is an unknown number of array sites (discussed further below). 

Using the values for the penetration depth at 900~nm and 450~nm in Table~\ref{table:penDepths}, we obtain an effective probe penetration depth of $\approx$~22~nm, which is $\approx$ 3.5 times smaller than the penetration depth of the pump beam, $\delta(\omega_{pump})$. 
This ratio being significantly greater than one helps to explain why simulations neglecting penetration depth effects agree well with experiment; the probe samples from a much smaller volume of the material than that which is excited by the pump pulse. Thus, in the region sampled by the probe, the inhomogeneity due to the penetration depth of the pump is a small effect. 

\begin{figure*}
    \centering
    \includegraphics[scale=0.6]{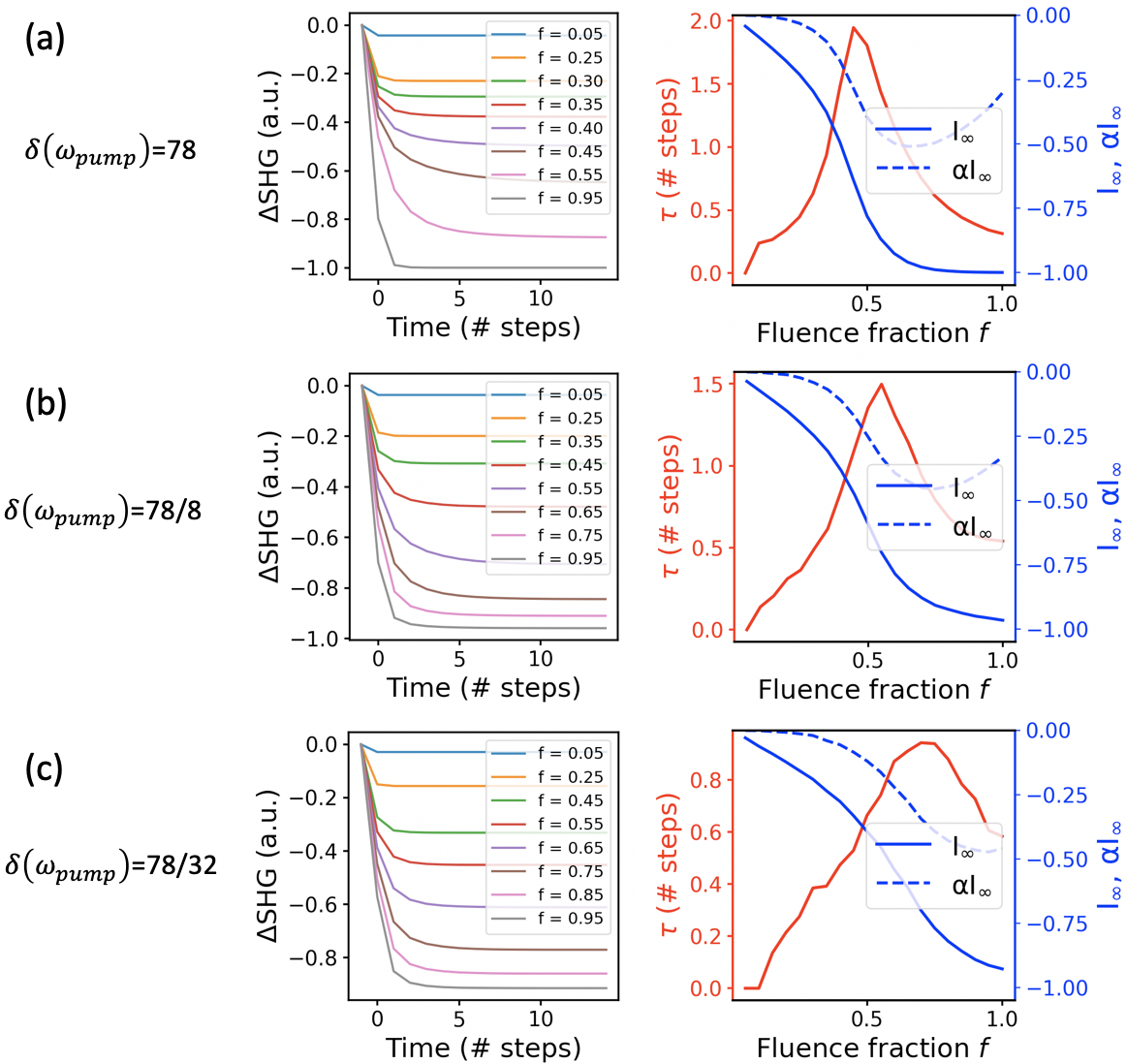}
    \caption{Simulation output accounting for the penetration depth of the pump $\delta(\omega_{pump})$ and of the probe $\delta^{eff}$. The ratio $\delta(\omega_{pump}) / \delta^{eff}$=3.5 is fixed in all cases. \textbf{(a-c)} show the results for $\delta(\omega_{pump})$ set to 78, 9.8, and 2.4 sites, respectively.}
    \label{fig:penDepthSims}
\end{figure*}

In incorporating finite penetration depth effects into the simulations, the ratio $\delta(\omega_{pump}) / \delta^{eff}$ must be preserved. However, we do not know the magnitude of the penetration depths that should be used in the simulations (in units of number of array sites) to obtain the closest correspondence to the physical experiment. Therefore, in Figure~\ref{fig:penDepthSims} we present the results of several simulations in which the magnitude of the penetration depths is varied while the ratio $\delta(\omega_{pump}) / \delta^{eff}$ = 3.5 is fixed. 
Simulation output corresponding to several different values of $\delta(\omega_{pump})$ are shown. 
We have estimated that each site in the simulations corresponds to a length on the order of $\sim$1~nm in the experiment (c.f. Section \ref{sec:simScales}). Thus, in Fig.~\ref{fig:penDepthSims}(a) we present the simulation output with $\delta(\omega_{pump})$=78~sites (and $\delta^{eff}$=22~sites). Because this value is larger than the size of our simulated array, we further include the results of simulations with smaller values for the penetration depth. 
In part due to the ratio $\delta(\omega_{pump}) / \delta^{eff}\gg$~1, the effects of the penetration depth on the fluence dependence of $\tau$, $I_{\infty}$, and $\alpha$ are small.

\section{Simulated Cluster Properties}

\begin{figure}
    \centering
    \includegraphics[scale=0.6]{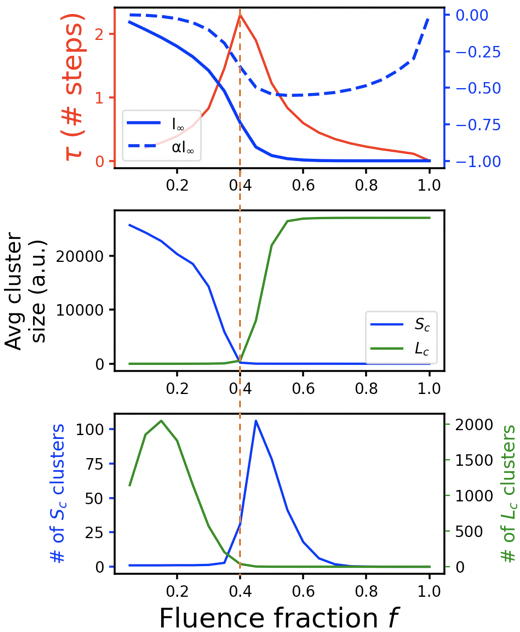}
    \caption{Cluster behavior in the 40x40x40 simulation with $\sigma_{th}$=3 and $\beta$=0. \textbf{(top)} Reproduction of Fig. 3(b) for comparison: $\tau$, $I_{\infty}$, and $\alpha I_\infty$ vs. fluence fraction $f$ \textbf{(middle)} Average area of the simulated $S_c$ and $L_c$ clusters at $t\xrightarrow{}\infty$ vs. fluence fraction \textbf{(bottom)} number of $S_c$ and $L_c$ clusters at $t\xrightarrow{}\infty$ in the simulations}
    \label{fig:stackClust}
\end{figure}

Better understanding of the PIPT in CRO and related compounds may come from future experimental investigations of the photoinduced clusters. 
In Fig. \ref{fig:stackClust} we present the number of clusters and the average cluster size of the $L_c$ and $S_c$ states found in our simulations once site-switching has stopped (using the same simulation settings as used in the main text, i.e. a 40x40x40 array with $\sigma_{th}$ = 3 and $\beta$ = 0).
A cluster here is defined as a collection of sites possessing the same state (i.e. $L_c$ or $S_c$) connected by an unbroken chain of nearest neighbor links. 

\begin{figure}
    \centering
    \includegraphics[scale=0.35]{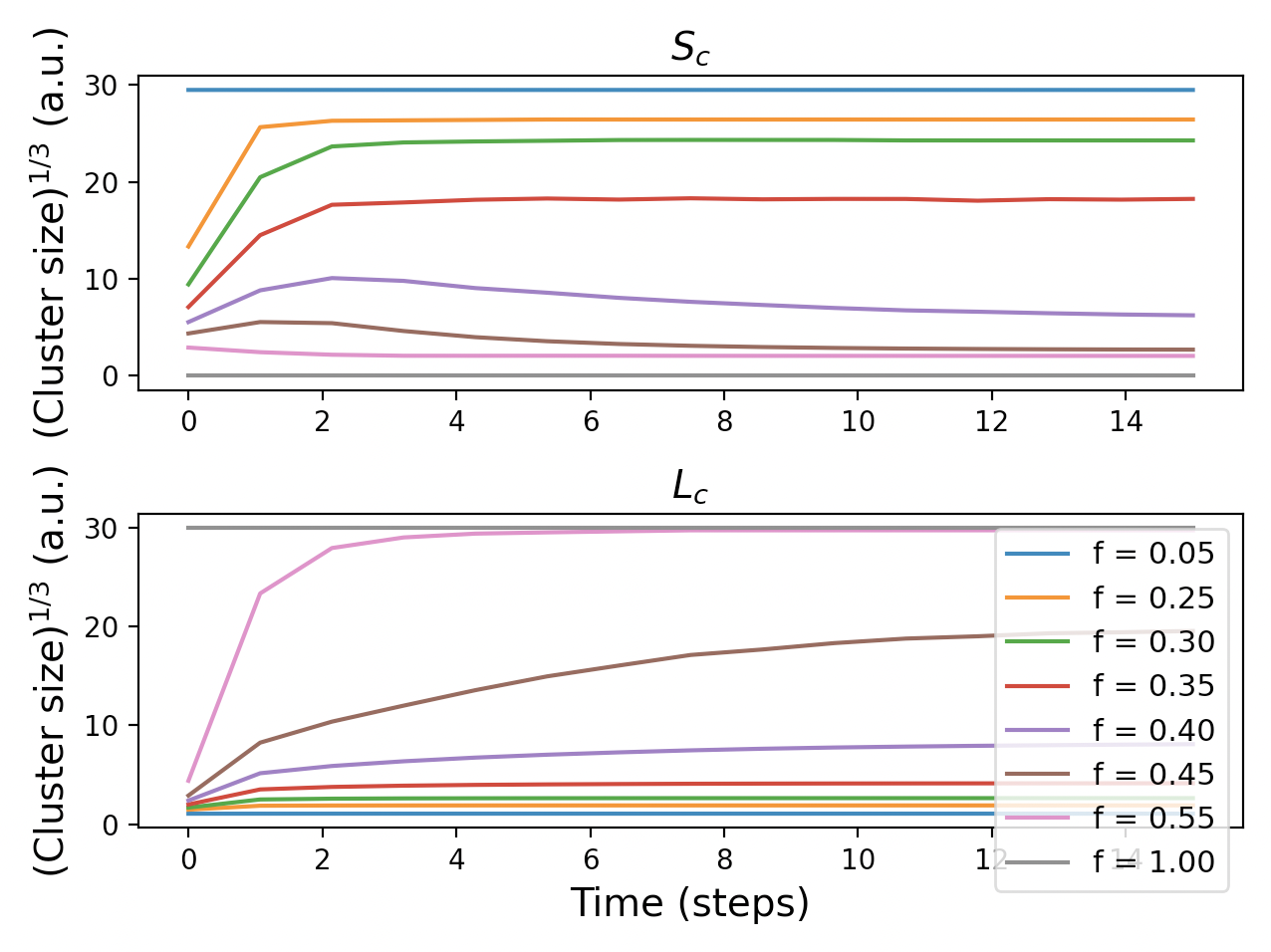}
    \caption{The cubed root of the average $L_c$ and $S_c$ cluster sizes vs. time steps in the $3d$ simulations with $\sigma_{th}$=3 and $\beta$=0.}
    \label{fig:ClSizeVsT}
\end{figure}

\begin{figure}
    \centering
    \includegraphics[scale=0.35]{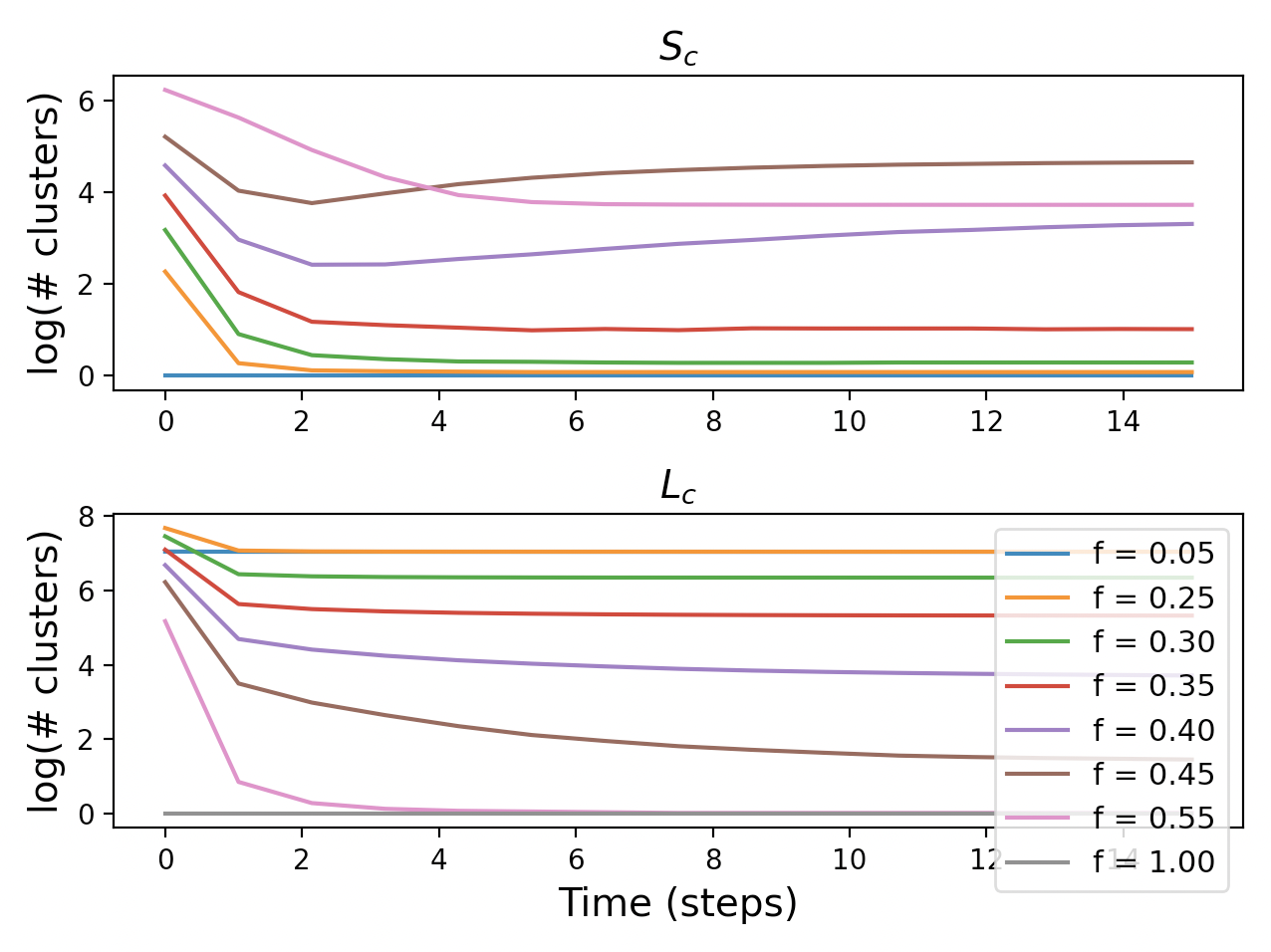}
    \caption{The logarithm of the average number of $L_c$ and $S_c$ clusters vs. time steps in the $3d$ simulations with $\sigma_{th}$=3 and $\beta$=0.}
    \label{fig:NumClustVsT}
\end{figure}

Figures \ref{fig:ClSizeVsT} and \ref{fig:NumClustVsT} show the time-dependence of the cluster sizes and the number of clusters, respectively, of both the $L_c$ and $S_c$ phases in the simulation presented in the main text (i.e. a 40x40x40 array with $\sigma_{th}$=3 and $\beta$=0).

The sigmoid-like dependence of the cluster size on the fluence, and the time dependence of the cluster sizes could be observed in future experiments using time-resolved diffuse x-ray or electron diffraction to verify our model of the photo-induced insulator-metal transition as a percolation phenomenon.

\section{Temperature Effects}
\begin{figure}[h!]
    \centering
    \includegraphics[scale=0.5]{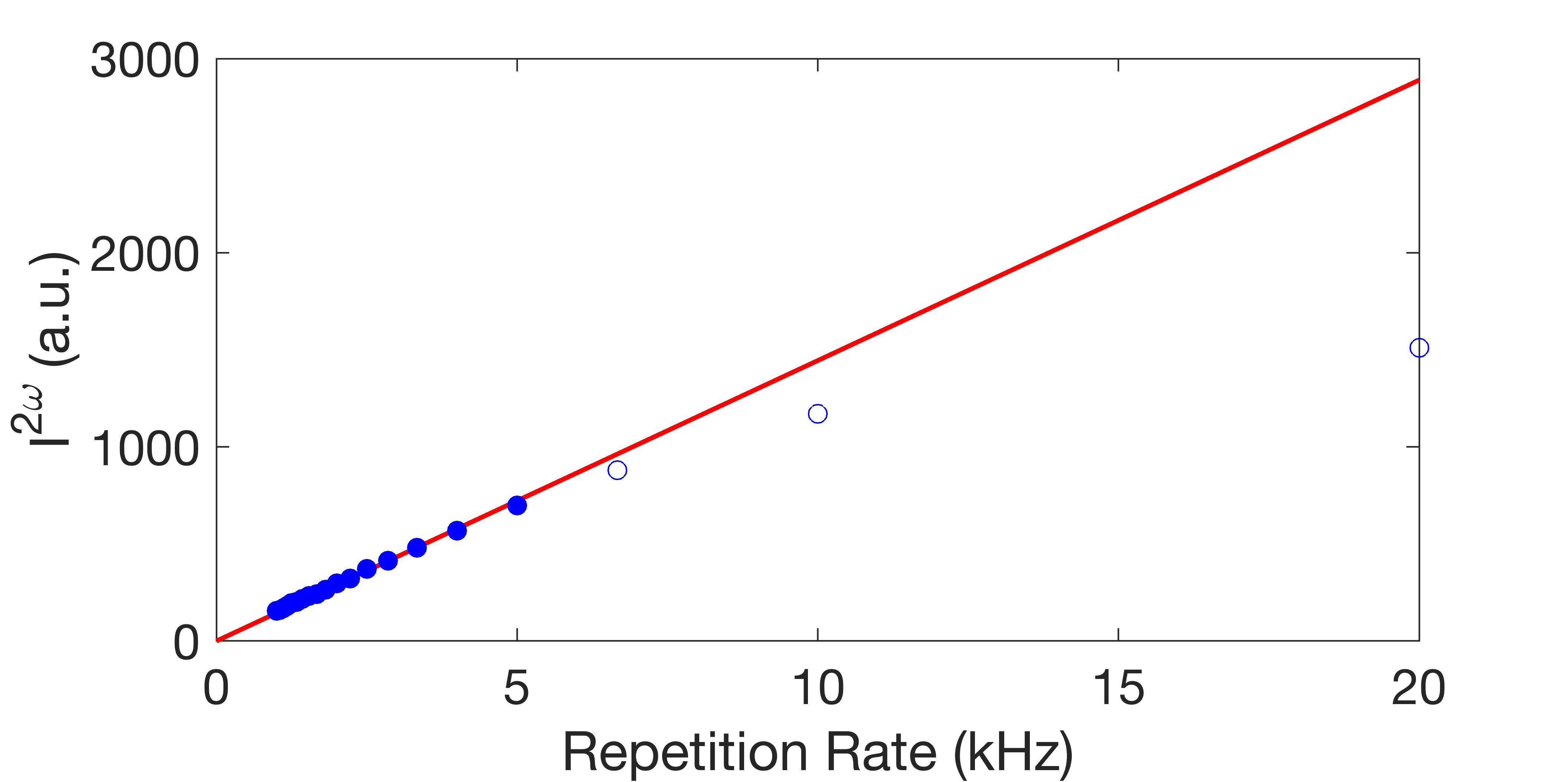}
    \caption{$I^{2\omega}$ shortly before the arrival of a 1.2 mJ/$\mathrm{cm}^2$ pump pulse vs. repetition rate of the pump and probe pulses. The red line is a linear fit (with zero offset) to the filled data points. The data included in the text was taken at 5 kHz.}
    \label{FigRRT}
\end{figure}

To ensure that the sample fully recovered between pulses, we varied the repetition rate of the laser (both pump and probe pulses) and fixed the probe pulse to arrive shortly before the pump pulse, while keeping the pump and probe fluences fixed. 
If the sample fully recovers between pulses, $I^{2\omega}$ will exhibit a linear trend with the repetition rate (because the intensity is proportional to the number of probe pulses), while a sublinear trend indicates an incomplete recovery of the sample between pump pulses.
Figure \ref{FigRRT} shows $I^{2\omega}$ vs. the repetition rate for the largest pump fluence used in this experiment, 1.3 mJ/$\mathrm{cm}^2$. We see that at repetition rates above the 5 kHz used in the experiment, the sample no longer recovers between pulses of this fluence. This is verified by performing a linear fit (with zero offset) to the data points $\leq$ 5 kHz. 

\subsection{Transient pump-induced heating}
From the heat capacity, we can determine an upper bound on the transient increase in the lattice temperature following the pump pulse. To do this, we use the following expression:
\begin{equation}
    q=\int_{T_i}^{T_f}\mathit{dT}c(T),
\end{equation}
where $q$ is the average heat absorbed per unit cell from a single pulse, $T_{i(f)}$ is the initial(final) lattice temperature, and $c(T)$ is the heat capacity per unit cell. The heat capacity as a function of temperature is extracted from Ref. \cite{VARADARAJAN2007402}, which used a flux-grown sample (as was used in our measurements). Specifically, this allows us to extract coefficients for the $T$ and $T^3$ terms in $c(T)$.

The following equation gives the absorbed energy per unit cell $q$, calculated with a weighted average in which the weight is proportional to the probe sensitivity:
\begin{equation}
\begin{aligned}
    q&=(1-R)F \frac{V_{uc}}{\delta(\omega_{pump})} \frac{\int_0^\infty dz e^{-z/\delta(\omega_{pump})}e^{-z/\delta^{eff}}}{\int_0^\infty dz e^{-z/\delta(\omega_{pump})}} \\
    &= (1-R)F \frac{V_{uc}}{\delta(\omega_{pump})} \frac{1}{1+\frac{\delta^{eff}}{\delta(\omega_{pump})}},
\end{aligned}
\end{equation}
where $\delta(\omega)$ = 78~nm is the penetration depth of the pump beam, $\delta^{eff}$ = 22~nm is the effective penetration depth of the SHG probe (c.f. Eq.~\ref{eq:effPenDepth}), $F$ is the fluence of the pump pulse, and $R$ = .18 is the reflectivity at the pump wavelength.

The upper bound on the transient temperature $T_f$ following the pump pulse is determined by equating these two expressions for the absorbed energy per unit cell $q$. 

The only remaining parameter in this calculation is the crystal temperature before the arrival of the pump pulse $T_i$. Without laser heating, the crystal is at a nominal temperature of $\approx$4 K. From Fig. 1(a), we see that the steady-state temperature increase due to the probe laser is $\approx$2 K near the transition temperature $T_{MI}$. However, at lower temperatures the heat capacity is smaller and the steady-state temperature increase due to the probe beam will be larger. But, because we are interested in obtaining an upper bound for the transient temperature increase, it suffices to assume a large initial temperature due to the heating from the probe laser. Table \ref{table:pumpheating} presents the obtained upper bound on the transient lattice temperature $T_f$ immediately following the pump pulse for various initial temperatures and for fluences $F_{sat}$ = 0.4 mJ/cm$^2$ and $F_{max}$ = 1.2 mJ/cm$^2$. 

\begin{table}[h]
\centering
    \begin{tabular}{|c||c|c|}
    \hline
      $T_i$ (K)  & $T_f$ @ $F=F_{sat}$ (K) & $T_f$ @ $F=F_{max}$ (K)\\
      \hline
      10  & 31 & 41 \\
      \hline
      20 & 32 & 42 \\ 
      \hline 
      25 & 34 & 43 \\
      \hline 
    \end{tabular}
    \caption{Upper bounds on the lattice temperature $T_f$ immediately following the arrival of the pump laser pulse for pump fluences $F_{sat}$ = 0.4 mJ/cm$^2$ and $F_{max}$ = 1.2 mJ/cm$^2$ and several initial steady-state temperatures $T_i$. }
    \label{table:pumpheating}
\end{table}

\subsection{Temperature-dependent time traces} \label{SuppTempTraces}

\begin{figure*}[h]
    \centering
    \includegraphics[scale=0.7]{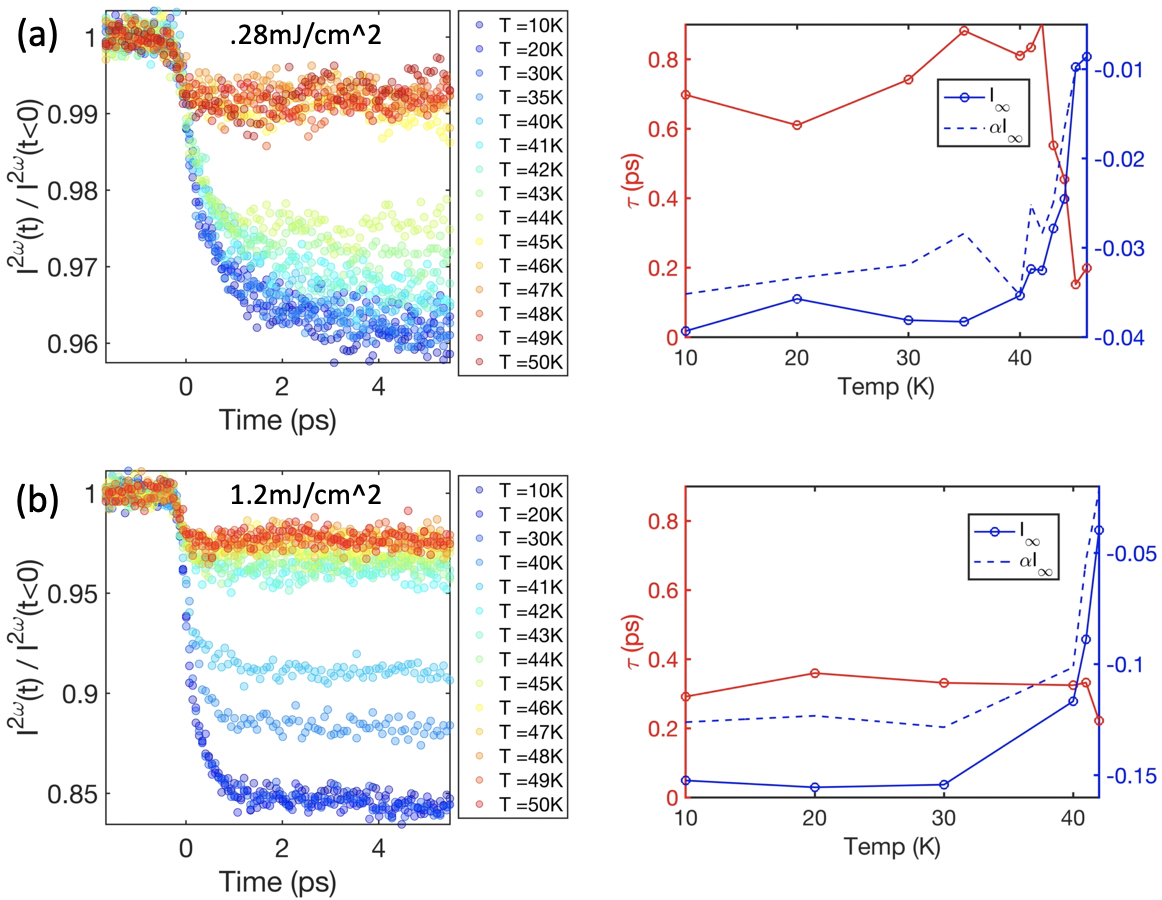}
    \caption{SHG time traces at varying temperature and the corresponding fit parameters extracted from fitting to Eq. 1. \textbf{(a)} The pump fluence is fixed at 0.28 mJ/$\mathrm{cm}^2$ and the polarization is such that $I^{2\omega}\propto|\chi^{ED}_{baa}|^2$. \textbf{(b)} The pump fluence is 1.2 mJ/$\mathrm{cm}^2$ and the polarization is such that $I^{2\omega}\propto|\chi^{ED}_{bbb}|^2$. The two datasets were taken at different locations on the sample.}
    \label{FigTemptraces}
\end{figure*}

The complete data set corresponding to Fig. 2(b) with pump fluence 0.28 mJ/$\mathrm{cm}^2$ is shown in Figure \ref{FigTemptraces}(b). Comparison with the data shown in \ref{FigTemptraces}(a) serves to highlight the variation in these measurements. 
Fig. \ref{FigTemptraces}(a) shows a very long relaxation timescale at temperatures just below the transition. However, no similar trend appears in Fig. \ref{FigTemptraces}(b), where time traces at temperatures near the transition are flat after several ps and have values between those at temperatures far above and below the transition. 
As mentioned previously, the domain structure along the $c$ axis has an effect on the measured SHG~\cite{Lei}, and this may be related to the differences between the datasets presented here. 

\begin{widetext}
\section{Landau theory for SHG near the metal-insulator transition}\label{sec:Landau theory for SHG near the metal-insulator transition}

We develop a Landau theory for the metal-insulator transition at $T_{MI}$=48~K.  (Laser heating is ignored in this discussion.)  Our main purpose is to calculate the in-plane components of the SHG susceptibility tensor, $\chi^{ED}_{ijk}\equiv\chi_{ijk}$ (where $i,j,k$ take values in $\{a,b\}$), for $T$ near $T_{MI}$. 

To describe the transition itself, independent of its relation to SHG, we postulate that a symmetry-preserving order parameter $\eta$ undergoes a first-order phase transition at $T=T_{MI}$.  In this part of the calculation, we follow Ref. \cite{KuznetsovHightemperature2003}, in which a Landau theory is developed for a symmetry-preserving order parameter (see also Ref. \cite{AzzolinaLandau2020}).

The symmetry-preserving order parameter in our calculation is intended to describe the change in crystal structure at the transition and could be, for instance, the volume change of the unit cell.  The microscopic nature of this order parameter does not affect the calculations below.  Indeed, the calculation below seems to be equally applicable if the order parameter is instead taken to describe the change in electronic structure, since this is also symmetry-preserving.  However, as discussed in the main text, experimental evidence from the Fe-doped compound disfavors either the electronic or magnetic structural change as explanations for the jump in SHG.

To describe SHG, we provide two phenomenological approaches, neither of which we attempt to justify microscopically, but both of which yield the same phenomenological parameterization of the measured quantity $I^{2\omega}$.  In the first approach, we expand the SHG susceptibility in the order parameter $\eta$.  Here we have taken inspiration from the Landau calculation of elastoresistivity in Ref. \cite{ShapiroSymmetry2015}.  In the second approach, we use a phenomenological time-averaged Landau free energy that includes frequency dependence, following and extending the approach of Refs. \cite{PershanNonlinear1963,Sageneralized2000}.  Ref. \cite{Sageneralized2000} has developed this approach for the case of a symmetry-breaking order parameter; in this case, certain SHG tensor components must vanish above the transition temperature by symmetry, but can be non-vanishing below the transition temperature because the order parameter reduces the symmetry.  In the case we consider, the same SHG tensor components are allowed both above and below $T_{MI}$ (since the transition does not break any symmetries), and we describe this in an approach that goes beyond that of Ref. \cite{Sageneralized2000} in that we explicitly minimize over frequency-dependent internal degrees of freedom in the time-averaged Landau free energy.

We organize the calculation as follows.  First, we review the symmetry-preserving transition as developed by Ref.~\cite{KuznetsovHightemperature2003}.  Second, we discuss how symmetries constrain the tensor elements of the SHG susceptibility.  Then we present two phenomenological approaches to obtaining the jump in $I^{2\omega}$.

\subsection{Symmetry-preserving transition}\label{sec:Symmetry-preserving transition}
We recall here the Landau theory provided by Kuznetsov et al. \cite{KuznetsovHightemperature2003} for a generic symmetry-preserving phase transition.  Taking $\eta$ to be a symmetry-preserving order parameter, one writes
\beq
    \mathcal{F}_0 = a_1 \eta + a_2 \eta^2 + a_3 \eta^3 + \eta^4.\label{eq:F_0}
\eeq
The cubic term is eliminated by the change of variables $\Delta = \eta + a_3/4$,  $\tilde{a}_1 = a_1 + a_3^3/8-a_2 a_3/2$,  $\tilde{a}_2 = a_2 - \frac{3}{8}a_3^2$, yielding 
$\mathcal{F}_0 = \tilde{a}_1 \Delta + \tilde{a}_2\Delta^2 + \Delta^4$, where a $\Delta$-independent constant has been dropped.  The solutions to $\partial \mathcal{F}_0 /\partial \Delta=0$ are presented in Ref. \cite{KuznetsovHightemperature2003}.  The final result that we need for the present calculation is that if $D\equiv 8\tilde{a}_2^3 + 27\tilde{a}_1^2 <0$ (note that this requires $\tilde{a}_2<0$), then $F_3$ is an asymmetric double well, with the global minimum being on the left if $\tilde{a_1} > 0$ and on the right if $\tilde{a_1}<0$.  Ref. \cite{KuznetsovHightemperature2003} presents three solutions (in the case $D<0$) called $\Delta_1$, $\Delta_2$, and $\Delta_3$.  We find (contrary to Ref. \cite{KuznetsovHightemperature2003}) that the global minimum is $\Delta_1$, i.e., the solution defined by 
\beq
    \Delta_1 \equiv - 2 \rho \cos \frac{\varphi}{3},
\eeq
where $\rho = \sgn \tilde{a}_1 \sqrt{|\tilde{a}_2|/6}$ and $\varphi = \arccos \left(\frac{\tilde{a}_1}{8\rho^3}\right)$ (note that $\varphi$ is real because of $D<0$).  
\begin{figure}
    \centering
    \includegraphics[width=0.5\linewidth]{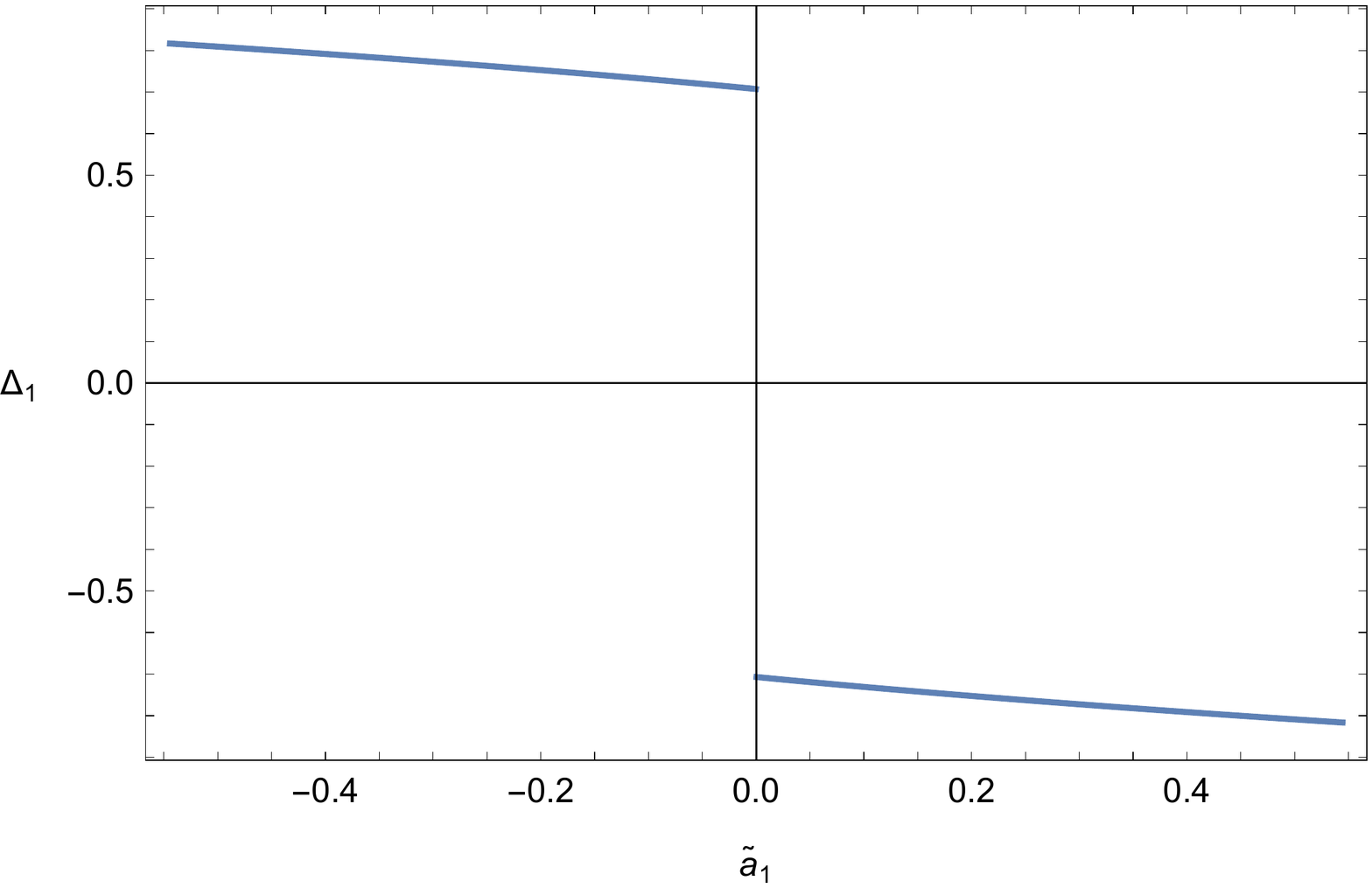}
    \caption{The global minimum $\Delta_1$ for $\tilde{a}_2=-1$ (all units are arbitrary).}\label{fig:Delta1}
\end{figure}
To get a first order phase transition for our problem, we fix $\tilde{a}_2<0$ and assume a linear dependence on temperature near $T_{MI}$:
\beq
    \tilde{a}_1 = \tilde{a}_{1,0}(T-T_{MI}),\label{eq:tildea1}
\eeq
where $\tilde{a}_{1,0}$ is a temperature-independent constant.  Then $\Delta_1$ (and thus also the equilibrium value of the original order parameter $\eta$) jumps discontinuously at $T=T_{MI}$.

For later convenience, we write the global minimum in an alternate form with the temperature dependence explicitly included:
\beq
    \Delta_1(T) =  - \sqrt{\frac{|\tilde{a}_2|}{2}} \left[ \sgn\bpl\tilde{a}_{1,0}(T-T_{MI})\bpr  + f\bpl\tilde{a}_{1,0}(T-T_{MI})\bpr \right],\label{eq:Delta1(T) exact}  
\eeq
where $f(x)$ is a smooth function defined by
\beq
    f(x) =
    \begin{cases}
        \frac{2}{\sqrt{3}}\cos \left( \frac{1}{3}\arccos\left[ \frac{1}{8}\left( \frac{|\tilde{a}_2|}{6}\right)^{-3/2} x \right] \right)- 1 & 0 <  x < (\frac{2}{3}|\tilde{a}_2|)^{3/2},\\
        -f(-x) & -(\frac{2}{3}|\tilde{a}_2|)^{3/2}<x< 0.
    \end{cases}\label{eq:f}
\eeq
We note in particular the expansion of $\Delta_1(T)$ near $T=T_{MI}$:
\beq
    \Delta_1(T) =  - \sqrt{\frac{|\tilde{a}_2|}{2}} \left[ \sgn\bpl\tilde{a}_{1,0}(T-T_{MI})\bpr  +\frac{\tilde{a}_{1,0}(T-T_{MI})}{2\sqrt{2|\tilde{a}_2|^3}}\right] + O((T-T_{MI})^2).\label{eq:Delta1(T) expanded}   
\eeq
Thus, at the transition temperature, $\Delta_1(T)$ jumps discontinuously with a continuous slope whose sign at $T=T_{MI}$ is the same as the sign of the discontinuity [i.e., $\Delta_1(T)$ resembles either Fig. \ref{fig:Delta1} or the same figure flipped vertically].  The global minimum $\eta_0(T)$ of the original order parameter is then obtained by the additive shift $\eta_0(T)=\Delta_1(T) -a_3/4$.

From Eq. \eqref{eq:Delta1(T) expanded}, we see that, in order to obtain a consistent expansion to linear order in $T-T_{MI}$, it may be necessary to include temperature dependence in some of the coefficients ($a_2$ and $a_3$) that we took as constant in Eq. \eqref{eq:F_0}, i.e., set $\tilde{a}_2 = \tilde{a}_{2,0}+ \tilde{a}_{2,1}(T- T_{MI}) 
+ O((T- T_{MI})^2)$ (with $\tilde{a}_2<0$ always) and $a_3 = a_{3,0} + a_{3,1}(T-T_{MI})+ O((T-T_{MI})^2)$.  Eq. \eqref{eq:Delta1(T) expanded} then would yield
\beq
    \eta_0(T) = A + B\sgn(T-T_{MI}) + C (T-T_{MI}) + D |T-T_{MI}| + O((T-T_{MI})^2),\label{eq:eta_0 final expansion}
\eeq
where $A,B,C,D$ are temperature-independent constants [e.g., $A = -\sqrt{|\tilde{a}_{2,0}|/2}\sgn \tilde{a}_{1,0}$].  In other words, $\eta_0(T)$ jumps discontinuously and its slope can be different on either side of the jump.

For the expansion \eqref{eq:F_0} to make sense, the order parameter must be small [e.g., we should have $|B|\ll |A|$ in \eqref{eq:eta_0 final expansion}].  A strongly first order transition would not necessarily fit into this framework.

\subsection{Symmetry-allowed components of the SHG susceptibility tensor}\label{sec:Symmetry-allowed components of the SHG susceptibility tensor}
Next, we recall how inversion symmetry along a particular axis is broken in CRO through the scenario of hybrid improper ferroelectricity \cite{BenedekHybrid2011,HarrisSymmetry2011,Benedek2012,NowadnickDomains2016}.  It is convenient to consider the components of the polar moment along the crystallographic axes $\hat{\mathbf{a}}\equiv [110]$ and $\hat{\mathbf{b}}\equiv[\bar{1}10]$:
\beq
    P_a \equiv \mathbf{P}\cdot\hat{\mathbf{a}},\  P_b\equiv\mathbf{P}\cdot\hat{\mathbf{b}}.\label{eq:Pa and Pb}
\eeq

In the following discussion, we use the notation and results of Harris (Ref. \cite{HarrisSymmetry2011}).  While Ref. \cite{HarrisSymmetry2011} focuses on $\text{Ca}_3\text{Mn}_2\text{O}_7$ and $\text{Ca}_3\text{Ti}_2\text{O}_7$, we can expect CRO to fit into the same framework.

The key point is that the full symmetry of the reference tetragonal structure $I4/mmm$ is broken down to $Cmc2_1$ ($\#36$) by the condensation of an octahedral tilt mode and an octahedral rotation mode.  [We note here that $Cmc2_1$ is an alternative expression for the space group $Bb2_1m$ mentioned in the main text; see footnote 13 in Ref. \cite{Yoshida05}.]  Furthermore, there is a symmetry-allowed trilinear coupling between the tilt mode, rotation mode, and polar moment.  There are 8 possible domains: first there are two choices for the wavevector $\mathbf{q}$ associated with both the tilt and rotation mode, and then there four choices corresponding to the tilt mode and rotation mode each have an independently chosen sign.  The spontaneous polar moment lies either along the $a$ axis or $b$ axis with either sign, so the $8$ domains correspond to four possible directions for the spontaneous polar moment, i.e., $90$ degree domains.  (Recall that we focused in the experiment on a region of the sample in which any domains present, whether one or more than one, were aligned along a single axis.  However, $90$ degree domains were observed in other samples.)  We note that, up to an overall constant (which is of no concern since we only wish to establish the axis along which inversion is broken), we may identify the $x$ and $y$ components of $\mathbf{P}$ with $Q_{5,1}$ and $Q_{5,2}$ from Ref. \cite{HarrisSymmetry2011}.  Thus, again up to an overall constant, our $P_a$ and $P_b$ are $Z_5$ and $-Y_5$ from Ref. \cite{HarrisSymmetry2011}.  See Eq. (7) of Ref. \cite{HarrisSymmetry2011} for the trilinear coupling.

The condensation of these order parameters described above occurs at some temperature much higher than the transition $T_{MI}$ that is our focus.  For our purpose of studying $T$ near $T_{MI}$, we only need to note that inversion is broken along the axis of the spontaneous polar moment (either $a$ or $b$, depending on the domain).  The symmetries that mix $a$ and $b$ components together (see Table II and Eq. (2) of Ref. \cite{HarrisSymmetry2011}) are also broken.  We thus see that the symmetry-allowed tensor elements of the SHG susceptibility are as in \eqref{eq:chi allowed}.

From here on, we focus on the three tensor elements relevant to the experimental geometry, namely, $\chi_{aba},\chi_{baa}$, and $\chi_{bbb}$ (note that $\chi_{aab}=\chi_{aba}$).  Also, we consider only the case of a uniform sample (no domains).

\subsection{First approach: direct expansion}\label{sec:First approach: direct expansion}
For the regime of $T$ near $T_{MI}$ that is our focus, we can obtain a phenomenological expression for the SHG susceptibility by supposing that it is analytic in the symmetry-preserving order parameter $\eta$. 
In particular, we write
\beq
    \chi_{ijk} = A^{(\omega)}_{ijk} + B^{(\omega)}_{ijk} \eta + O(\eta^2),\label{eq:chi expand in eta}
\eeq
where $(ijk)=(aba),(baa),$ or $(bbb)$ and where $A^{(\omega)}_{ijk}, B^{(\omega)}_{ijk}$ are constants (possibly complex) that could depend on temperature as well as frequency.  This approach has some similarity to the Landau theory calculation of elastoresitivity in Ref. \cite{ShapiroSymmetry2015}, though the symmetry considerations here are different.

It is simplest to assume that $A^{(\omega)}_{ijk}, B^{(\omega)}_{ijk}$ are smooth functions of temperature (at least for $T$ near $T_{MI}$).  Then, the discontinuous jump in the equilibrium value $\eta_0(T)$ as $T$ crosses $T_{MI}$ (see Sec. \ref{sec:Symmetry-preserving transition}) yields a corresponding jump in each $\chi_{ijk}$ (weighted by the appropriate coefficient $B^{(\omega)}_{ijk}$ evaluated at $T=T_{MI}$).  In this way, we have related a first-order, symmetry-preserving phase transition with a jump in SHG susceptibility.

\subsection{Second approach: Time-averaged Landau free energy}\label{sec:Second approach: Time-averaged Landau free energy}
We now present another phenomenological calculation of the SHG susceptibilty for $T$ near $T_{MI}$.  Here, the assumption \eqref{eq:chi expand in eta} is replaced by two other assumptions: (a) neglect of dissipation and (b) the existence of a time-averaged Landau free energy (see below).  This approach yields the same phenomenological parameterization (of $\chi_{ijl}$) as the approach of Sec. \ref{sec:First approach: direct expansion} except with real coefficients instead of complex, due to the assumption (a).

In Ref. \cite{PershanNonlinear1963}, Pershan proposed a phenomenological ``time-averaged free energy'' function that includes frequency dependence and can be used to calculate nonlinear optical effects.  In Ref. \cite{Sageneralized2000}, Sa et al. extended this approach further by including coupling to an order parameter in the time-averaged free energy.  The time-averaged free energy as considered by Ref. \cite{Sageneralized2000} is a frequency-dependent analogue to the usual (frequency-independent) equilibrium free energy; it is sufficient for determining whether or not a particular tensor component of the SHG susceptibility is or is not constrained to be zero above a symmetry-breaking transition temperature.  Since in our present problem we are considering a transition that does not break any symmetries, we must develop the approach of Ref. \cite{Sageneralized2000} further in order to calculate the change in tensor components that are symmetry-allowed both above and below the transition.  To do this, we use a frequency-dependent analogue to the usual (frequency-independent) Landau free energy.  The essential difference with \cite{Sageneralized2000}, then, is that we explicitly minimize over frequency-dependent degrees of freedom of the system.

\subsubsection{SHG in a toy model}\label{sec:SHG in a toy model}
As a warm-up, we consider a toy model of SHG (with no temperature dependence or phase transition): a one-dimensional model without inversion symmetry.  The sample is subject to external electric fields at frequencies $\omega$ and $2\omega$.  We write the amplitudes of the AC electric field and polar moment at frequency $\omega'$ ($\omega'=\omega$ or $2\omega$) as $E^{(w')}$ and $P^{(w')}$.  For a more careful discussion of the order of limits used in defining these Fourier amplitudes, see Eq. (2.8) and below in Ref. \cite{PershanNonlinear1963}.

The AC polar moments, which are the internal variables of the system, appear in all terms allowed by symmetry.  We have already accounted for translation symmetry by writing the electric fields and polar moments as position-independent quantities.  The only remaining symmetry of this toy model is time translation.  Under time translation by $t'$, $P^{(\omega')}\to e^{i t' \omega'}P^{(\omega)}$ and $P^{(\omega')*}\to e^{-i t' \omega'}P^{(\omega)}$, which implies that in any given term, the sum of frequencies in unconjugated polar moments must equal the sum of frequencies in conjugated polar moments.  For the time-averaged Landau free energy, we thus write the most general polynomial (up to fourth order) that satisfies time translation invariance:
\beq
    \mathcal{F}_\text{AC}^{(\text{toy})} = \sum_{\omega'=\omega,2\omega} (a^{(\omega')} |P^{(\omega')}|^2 + \frac{1}{2} b^{(\omega')} |P^{(\omega')}|^4 ) +  c^{(\omega)} |P^{(\omega)}|^2|P^{(2\omega)}|^2\\
    + 2\text{Re}[ d^{(\omega)}  P^{(2\omega)*} (P^{(\omega)})^2),\label{eq:FAC toy model}
\eeq
in which $a^{(\omega')}$, $b^{(\omega')}$, $c^{(\omega)}$, and $d^{(\omega)}$ are constants.  Although a complex $d^{(w)}$ is consistent with symmetry and with $\mathcal{F}_\text{AC}^{(\text{toy})}$ being real, an imaginary part seems to result either in dissipation (contrary to the basic assumption of the approach \cite{PershanNonlinear1963}) or in unphysical effects, so hereafter we assume that $d^{(\omega)}$ is real.  On physical grounds, we also assume that the parameters in $\mathcal{F}^{(\text{toy})}$ are such that the AC polar moment vanishes in the absence of an external electric field; for instance, $a^{(\omega')}$, $b^{(\omega')}$, and $c^{(\omega)}$ should all be positive.  

Including the coupling between the polar moment and the external electric field, we obtain:
\beq
    \mathcal{F}^{(\text{toy})} = \mathcal{F}_{\text{AC}}^{(\text{toy})} + \mathcal{F}_{\text{ext}}, 
\eeq
where
\beq
    \mathcal{F}_\text{ext}= \sum_{\omega'=\omega,2\omega} \left( -  E^{(\omega')}P^{(\omega)'*} -  E^{(\omega')*}P^{(\omega)'} \right).
\eeq
The minimum of $\mathcal{F}^{(\text{toy})}$ (varying the real and complex parts of $P^{(\omega)}$ and $P^{(2\omega)}$) is then a function of $(E^{(\omega)},E^{(2\omega)})$ only and coincides with the time-averaged free energy considered by Ref. \cite{Sageneralized2000}.

We proceed to calculate the linear susceptibility and SHG susceptibility in this toy model.  It is convenient to minimize in $P^{(\omega)}$ and $P^{(2\omega)}$ with $P^{(\omega)*}$ and $P^{(2\omega)*}$ treated as independent variables.  (This is equivalent to minimizing in the real and complex parts of $P^{(\omega)}$ and $P^{(2\omega)}$.)  To present the calculation more compactly, we define $4$-component arrays:
\bseq
\begin{align}
    \underline{P}&= (P^{(\omega)},P^{(\omega)*},P^{(2\omega)},P^{(2\omega)*}), \\
    \underline{E}&= (E^{(\omega)},E^{(\omega)*},E^{(2\omega)},E^{(2\omega)*}).
\end{align}
\eseq
The minimization condition may be written as $\partial\mathcal{F}^{(\text{toy})}/\partial \underline{P}^* =0 $, i.e., 
\beq
    \underline{E} = \frac{\partial\mathcal{F}_{\text{AC}}^{(\text{toy})}}{\partial \underline{P}^*}.\label{eq:E field toy model}
\eeq
The polar moment that minimizes $\mathcal{F}^{(\text{toy})}$ may generally be expanded
\beq
    \underline{P}_\alpha =  (\chi_1)_{\alpha\beta} \underline{E}_\beta + (\chi_2)_{\alpha\beta\gamma}\underline{E}_\beta\underline{E}_\gamma + \dots,
\eeq
where repeated indices are summed (from $1$ to $4$) and where $\underline{P}^s$ is the spontaneous polar moment and $\chi_1$,$\chi_2$ are the first- and second-order susceptibility tensors.  The SHG susceptibility is a particular component of $\chi_2$:
\beq
    \chi_\text{SHG} = \frac{1}{2}  \frac{\partial^2 P^{(2\omega)}  }{\partial (E^{(\omega)})^2 }\biggr\rvert_{\underline{E}=\underline{0}} = (\chi_2)_{311}.
\eeq

To calculate the susceptibilities, we write the $4$-by-$4$ matrix $\partial \underline{P}_\alpha / \partial \underline{E}_\beta$ as $\partial \underline{P} / \partial \underline{E}$ and use a simple matrix identity:
\beq
    \frac{\partial \underline{P}}{\partial\underline{E}} = \left( \frac{\partial \underline{E}}{\partial\underline{P}} \right)^{-1}.\label{eq:matrix inverse}
\eeq
Then we get $\chi_1 = (\partial \underline{E} / \partial \underline{P})^{-1}\rvert_{\underline{P}=\underline{0}} = \text{diag}(\chi_1^{(\omega)}, \chi_1^{(\omega)},\chi_1^{(2\omega)},\chi_1^{(2\omega)})$, where $\chi_1^{(\omega')} = 1/a^{(\omega')}$ [c.f. Eqs. \eqref{eq:FAC toy model} and \eqref{eq:E field toy model}].

To get the second-order susceptibility, we expand $\partial \underline{E} / \partial \underline{P}$ to another order:
\beq
   \frac{\partial \underline{E}}{\partial \underline{P}} = \chi_1^{-1} + \frac{\partial^2 \underline{E}}{\partial \underline{P}_\alpha \partial\underline{P}}\underline{P}_\alpha + \dots, 
\eeq
where the omitted terms are quadratic or higher in $\underline{P}$.  Then
\beq
    \left(\frac{\partial \underline{E}}{\partial \underline{P}}\right)^{-1} = \chi_1 -\chi_1 \frac{\partial^2 \underline{E}}{\partial \underline{P}_\alpha \partial\underline{P}} \chi_1 \underline{P}_\alpha + \dots,
\eeq
which yields the SHG susceptibility:
\bseq
\begin{align}
    \chi_\text{SHG}&= \frac{1}{2} \left\{ \frac{\partial \underline{P}_\alpha}{\partial E^{(\omega)}} \frac{\partial}{\partial \underline{P}_\alpha }\left[ \left( \frac{\partial \underline{E}}{\partial \underline{P}} \right)^{-1} \right]_{31}  \right\}\biggr\rvert_{\underline{P}=\underline{0}}\\
    &= \frac{1}{2}  \left\{\chi_1^{(w)} \frac{\partial}{\partial P^{(\omega)}}\left[- \chi_1^{(2\omega)} \frac{\partial^2 E^{(2w)}}{\partial \underline{P}_\alpha \partial P^{(\omega)}} \chi_1^{(\omega)} \underline{P}_\alpha\right]\right\}  \biggr\rvert_{\underline{P}=\underline{0}}\\
    &= - \frac{1}{2}  \chi_1^{(2\omega)}(\chi_1^{(\omega)})^2 \frac{\partial^2 E^{(2w)}}{\partial^2 P^{(\omega)}}  \biggr\rvert_{\underline{P}=\underline{0}}\\
    &= - d^{(\omega)}\chi_1^{(2\omega)}(\chi_1^{(\omega)})^2 .
\end{align}
\eseq

\subsubsection{SHG near the metal-insulator transition in CRO}\label{sec:SHG near the metal-insulator transition in CRO}
Next, we combine the above ingredients into a time-averaged Landau free energy that describes the SHG susceptibility tensor in CRO near the transition at $T_{MI}$.

As our focus is on the in-plane (i.e., the $a$-$b$ plane) susceptibility, we ignore the $c$ axis.  We write the most general time-averaged Landau free energy consistent with inversion symmetry along $a$ (but broken inversion along $b$) and time translation invariance, keeping only up to quartic terms:
\beq
    \mathcal{F}= \mathcal{F}_0 + \mathcal{F}_{\text{AC}} + \mathcal{F}_\text{coupling} + \mathcal{F}_\text{ext},
\eeq
where $\mathcal{F}_0$ is the Landau free energy for the symmetry-preserving order parameter $\eta$ [Eq. \eqref{eq:F_0}], and where
\bseq
\begin{align}
    \mathcal{F}_\text{AC} &= \sum_{i=a,b}\left[ \sum_{\omega'=\omega,2\omega} \left( a_i^{(\omega')}|P_i^{(\omega')}|^2  + \frac{1}{2} b_i^{(\omega')}|P_i^{(\omega')}|^4\right) +c_{ij}^{(\omega)}|P_i^{(\omega)}|^2 |P_j^{(2\omega)}|^2\right]\notag\\
    &\qquad + 2\text{Re}[ 2d_{aba}^{(\omega)} P_a^{(2\omega)*}P_a^{(\omega)} P_b^{(\omega)} +d_{baa}^{(\omega)} P_b^{(2\omega)*}(P_a^{(\omega)})^2 + d_{bbb}^{(\omega)} P_b^{(2\omega)*} (P_b^{(\omega)})^2],\label{eq:FAC}\\
    \mathcal{F}_\text{coupling} &= \sum_{i=a,b} \sum_{\omega'=\omega,2\omega}\left(  \beta_i^{(\omega')} \eta |P_i^{(\omega')}|^2 + \lambda_i^{(\omega')} \eta^2 |P_i^{(\omega')}|^2 \right)\notag\\
    &\qquad +  2\text{Re}[ 2\delta_{aba}^{(\omega)}\eta P_a^{(2\omega)*}P_a^{(\omega)} P_b^{(\omega)} +\delta_{baa}^{(\omega)}\eta P_b^{(2\omega)*}(P_a^{(\omega)})^2 + \delta_{bbb}^{(\omega)}\eta P_b^{(2\omega)*} (P_b^{(\omega)})^2],\label{eq:Fcoupling}\\
    \mathcal{F}_{\text{ext}} &= \sum_{i=a,b}\sum_{\omega'=\omega,2\omega} 2\text{Re}[ -  P_i^{(\omega)'*}E_i^{(\omega')}].
\end{align}
\eseq
The $a$ and $b$ components of the polar moment and electric field are defined as in Eq. \eqref{eq:Pa and Pb}, now with frequency dependence.  The quantities $a_i^{(\omega')},b_i^{(\omega')},c_{ij}^{(\omega)}, d_{ijk}^{(\omega)}$ are all temperature-independent constants (as are $a_2$ and $a_3$ in $\mathcal{F}_0$).  The only temperature dependence in $\mathcal{F}$ is in the term $\tilde{a}_1$ that drives the first-order phase transition in $\eta$.  As in the toy model of the previous section, we assume that the constants are such that the spontaneous AC polar moment is zero.

The linear susceptibility and SHG susceptibilities are found similarly as in the toy model.  The arrays of polar moments and electric fields acquire an additional index specifying the component in the $a$-$b$ plane:
\bseq
\begin{align}
    \underline{P}&= (P_a^{(\omega)},P_b^{(\omega)},P_a^{(\omega)*},P_b^{(\omega)*},P_a^{(2\omega)},P_b^{(2\omega)},P_a^{(2\omega)*},P_b^{(2\omega)*}), \\
    \underline{E}&=(E_a^{(\omega)},E_b^{(\omega)},E_a^{(\omega)*},E_b^{(\omega)*},E_a^{(2\omega)},E_b^{(2\omega)},E_a^{(2\omega)*},E_b^{(2\omega)*}).
\end{align}
\eseq
The minimization conditions are $\partial \mathcal{F} / \partial \eta =0$ and $\partial \mathcal{F} / \partial \underline{P}_\alpha = 0$, the second of which yields the electric field as a function of polar moment:
\beq
    \underline{E}_\alpha = \frac{\partial (\mathcal{F}_\text{AC}+\mathcal{F}_\text{coupling})}{\partial \underline{P}_\alpha^*}.\label{eq:E field}
\eeq

We note that the $\eta$-dependent terms in $\mathcal{F}$ may be brought to the form
\beq
    \mathcal{F}_0 + \mathcal{F}_\text{coupling} = a_1(\underline{P})\eta + a_2(\underline{P}) \eta^2 + a_3 \eta^3 + \eta^4, 
\eeq
where
\bseq
\begin{align}
    a_1(\underline{P}) &= a_1 + \sum_{i=a,b} \sum_{\omega'=\omega,2\omega}  \beta_i^{(\omega')}  |P_i^{(\omega')}|^2 \notag\\
    &\qquad +  2\text{Re}[ \delta_{aba}^{(\omega)} P_a^{(2\omega)*}P_a^{(\omega)} P_b^{(\omega)} +\delta_{baa}^{(\omega)} P_b^{(2\omega)*}(P_a^{(\omega)})^2 + \delta_{bbb}^{(\omega)} P_b^{(2\omega)*} (P_b^{(\omega)})^2],\\
    a_2(\underline{P}) &= a_2 + \sum_{i=a,b} \sum_{\omega'=\omega,2\omega}  \lambda_i^{(\omega')}  |P_i^{(\omega')}|^2
\end{align}
\eseq
The minimization in $\eta$ thus yields a value $\eta_0(T,\underline{P})$ that can be obtained by replacing $a_1\to a_1(\underline{P})$ and $a_2\to a_2(\underline{P})$ in the solution found in Sec. \ref{sec:Symmetry-preserving transition}.  Note that $\eta_0(T,\underline{0})=\eta_0(T)$ [where $\eta_0(T)=\Delta_1(T)$ with $\Delta_1(T)$ given by Eqs. \eqref{eq:Delta1(T) exact}-\eqref{eq:Delta1(T) expanded}].  Also note that since $\underline{P}$ appears at least quadratically in $a_1(\underline{P})$ and $a_2(\underline{P})$, we have
\beq
    \frac{\partial \eta(\underline{P})}{\partial \underline{P}_\alpha}\biggr\rvert_{\underline{P}=\underline{0}} =0.\label{eq:eta first deriv zero}
\eeq

As in the toy model case, we calculate the first-order and SHG susceptibilities using Eq. \eqref{eq:matrix inverse} (which is now an $8$-by-$8$ matrix equation).  From $\chi_1= (\partial \underline{P} / \partial \underline{E})\rvert_{\underline{P}=\underline{0}}$, Eqs. \eqref{eq:FAC}-\eqref{eq:Fcoupling} and \eqref{eq:E field} yield
\beq
    \chi_1  \leftrightarrow \text{diag}(\chi_1^{(\omega)}, \chi_1^{(\omega)}, \chi_1^{(\omega)}, \chi_1^{(\omega)},\chi_1^{(2\omega)},\chi_1^{(2\omega)},\chi_1^{(2\omega)},\chi_1^{(2\omega)} ),
\eeq
where
\beq
    \chi_1^{(\omega')} = \frac{1}{a^{(\omega')} +\beta^{(\omega')}\eta_0(T) + \lambda^{(\omega')} \eta_0(T)^2 }.\label{eq:chi1}
\eeq
By the same manipulations as in the toy model case, we obtain the in-plane components of the SHG susceptibility tensor:
\beq
    \chi_{ijk} = - \left[d_{ijk}^{(\omega)} + \delta_{ijk}^{(\omega)}\eta_0(T)\right]\chi_1^{(2\omega)}(\chi_1^{(\omega)})^2,\label{eq:chiSHG}
\eeq
where $(ijk)=(aba),(baa),$ or $(bbb)$.

Dividing out the first-order susceptibilities yields:
\beq
    \frac{\chi_{ijk}}{\chi_1^{(2\omega)}(\chi_1^{(\omega)})^2} = A_{ijk}^{(\omega)}  + B_{ijk}^{(\omega)} \Delta_1(T),\label{eq:ratio}
\eeq
where $A_{ijk}^{(\omega)}$ and $B_{ijk}^{(\omega)}$ are temperature-independent constants determined by the various constants appearing in $\mathcal{F}$.  The temperature dependence of $\Delta_1(T)$ is discussed in Sec. \ref{sec:Symmetry-preserving transition}.

Furthermore, since in truncating $\mathcal{F}$ to quartic order we have already assumed $\eta$ to be a small parameter (note that this means $\tilde{a_2}$ must be small), we may expand Eq. \eqref{eq:chi1} in $\eta_0(T)$.  Collecting the various constants, we find that $\chi_{ijk}$ itself (even without dividing the first-order susceptibilities) also takes the form of the right-hand side of Eq. \eqref{eq:ratio} (with different constants $A_{ijk}^{(\omega)}$, $B_{ijk}^{(\omega)}$).

\end{widetext}

\bibliography{references}